\documentclass
[aps,pra,reprint,twocolumn,superscriptaddress,showpacs,showkeys,superscriptaddress,longbibliography]{revtex4-2}
\usepackage{dcolumn}    
\usepackage{ifpdf}
\usepackage{amssymb,lineno,amsfonts}
\usepackage{graphicx}   
\usepackage{dcolumn}    
\usepackage{bm}         
\usepackage{bbm}
\usepackage{mathrsfs}
\usepackage{upgreek}
\usepackage{mathtools}
\usepackage{epstopdf}
\usepackage{setspace}
\usepackage{hyperref}
\usepackage{array}
\usepackage{float}
\usepackage{amsmath}
\usepackage[normalem]{ulem}
\usepackage[usenames,dvipsnames]{xcolor}
\usepackage[matrix,frame,arrow]{xypic}
\usepackage[mathscr]{euscript}
\usepackage[dvipsnames]{xcolor}
\usepackage[]{qcircuit}

\usepackage{accents}
\newlength{\dhatheight}

\definecolor{med-blue}{RGB}{25,25,112}
\hypersetup{colorlinks, linkcolor={red},citecolor={blue}, urlcolor={MidnightBlue}}

\def\lket#1{\vert#1\rangle\hspace{-1mm}\rangle}

\newcommand{\ket}[1]{\vert{#1}\rangle}
\newcommand{\bra}[1]{\langle{#1}\vert}
\newcommand{\outpr}[2]{\vert{#1}\rangle\langle{#2}\vert}

\newcommand{\expv}[2]{\langle{#1}\rangle_{#2}}
\newcommand{\dmel}[2]{\langle{#1}\vert{#2}\vert{#1}\rangle}

\newcommand{\proj}[1]{\outpr{#1}{#1}}
\newcommand{\expec}[1]{\langle{#1}\rangle}

\begin{document}
\title{Observation of quantum phase-synchronization in a nuclear spin-system}
	
	\author{ V R Krithika}
\email{krithika\_vr@students.iiserpune.ac.in}
\affiliation{Department of Physics and NMR Research Center,\\
	Indian Institute of Science Education and Research, Pune 411008, India}
	\author{Parvinder Solanki}
\email{psolanki@phy.iitb.ac.in}
\affiliation{Department of Physics, Indian Institute of Technology-Bombay, Powai, Mumbai 400076, India}
	\author{Sai Vinjanampathy}
\email{sai@phy.iitb.ac.in}
	\affiliation{Department of Physics, Indian Institute of Technology-Bombay, Powai, Mumbai 400076, India}
	\affiliation{Centre for Quantum Technologies, National University of Singapore, 3 Science Drive 2, 117543 Singapore, Singapore}
	\author{T. S. Mahesh}
\email{mahesh.ts@iiserpune.ac.in}
\affiliation{Department of Physics and NMR Research Center,\\
	Indian Institute of Science Education and Research, Pune 411008, India}

	\begin{abstract}
{
We report an experimental study of phase-synchronization in a pair of interacting nuclear spins subjected to an external drive in nuclear magnetic resonance architecture. A weak transition-selective radio-frequency field applied on one of the spins is observed to cause  phase-localization, which is experimentally established by measuring the Husimi distribution function under various drive conditions. To this end, we have developed a general interferometric technique to directly extract values of the Husimi function via the transverse magnetization of the undriven nuclear spin. We further verify the robustness of synchronization to detuning in the system by studying the Arnold tongue behaviour.}
	\end{abstract}
		
\keywords{Quantum synchronization, NMR, phase-localization, Husimi distribution, Arnold tongue}

\maketitle

\section{Introduction}
\label{Introduction}
 Motivated by the stability and ubiquity of classical synchronization \cite{pikovsky2003synchronization}, quantum synchronization has been a field of intense study. Platforms that exhibit quantized dynamics have inspired theoretical studies in systems such as trapped ions \cite{lee2013quantum,hush2015spin}, superconducting circuits \cite{nigg2018observing}, atomic ensembles \cite{xu2014synchronization}, optomechanical systems \cite{ludwig2013prl,walter2014quantum,walter2015quantum,li2016quantum} and  nanomechanical systems \cite{zhang2012synchronization,holmes2012synchronization}. Theoretical studies have shown fundamental implications of synchronization to other fields such as entanglement generation \cite{timme2017classical,roulet2018quantum}, thermodynamics \cite{jaseem2020quantum,solanki2021role}, quantum networks \cite{li2017quantum} and continuous time crystals\cite{hajduvsek2021seeding}. 
 
Quantum synchronization proceeds by first considering the quantum analogue of phase space, which is one of many quasiprobability distrubitions such as Husimi function and Wigner function \cite{schleich2011quantum}. A valid limit cycle requires robustness against external perturbations and a neutral free phase \cite{pikovsky2003synchronization}. Thermal states were shown to be examples of such limit cycle states, showing equiprobable distribution of phases in phase space \cite{jaseem2020quantum}. Such phase space distribution functions have been used to construct measures of synchronization \cite{walter2014quantum,Galve2017,li2017properties,koppenhofer2019optimal,mari2013prl}. 
Following several studies of synchronization in specific systems, unified measures of quantum synchronization based on relative entropies have also been formulated \cite{jaseem2020generalized}. They used information theoretic measures based on quantum correlations, which have been shown to measure synchronization \cite{manzano2013synchronization,ameri2015mutual,roulet2018quantum}. These measures reduce to the phase space based measures under the system specific conditions \cite{jaseem2020generalized}. 

Following this theoretical interest \cite{walter2014quantum,walter2015quantum,bruder2018smallest,roulet2018quantum,sameer2018squuezing,jaseem2020generalized,jaseem2020quantum,solanki2021role,buca2021algebraic}, quantum synchronization has been experimentally demonstrated  recently in the IBM quantum computer \cite{koppenhofer2020quantum} and spin-1 cold atoms \cite{laskar2020observation}. 
The experimental demonstration of quantum synchronization is in general a challenging task owing to the difficulty in extracting required parameters. 
Quantum synchronization measurements typically require the system to settle into a steady state, which necessitates the need for long waiting times. Such long wait times allow for other experimental noise sources to interfere with the signal. Besides this, tomographic reconstruction of the state scales with the system size, requiring $O(n^2)$ measurements for a system of dimension $n$ \cite{bengtsson2006geometry}. Thus, the steady state characteristics of the system are inferred by extrapolating transient dynamics or by devising complicated measurement schemes.



Here, we resolve the aforementioned issues and report the observation of  synchronization in nuclear spins.
Nuclear magnetic resonance (NMR) architecture allows convenient control and manipulation of multi-spin systems using radio frequency (RF) pulses. These, in addition to inherent relaxation mechanisms, which cause the system to thermalize offer an ideal platform to observe synchronization.  Our experiment involves a pair of interacting spin-1/2 nuclei and a weak transition-selective RF field applied on one of the spins. The resulting phase-localization is characterized using Husimi distribution, which typically needs quantum state tomography (QST) \cite{krithika2019nmr}. 
However, QST after a long external drive proved to be highly inefficient because, as the steady state is approached, the irradiated levels saturate and relevant transitions become too faint.
We overcome this by introducing a new interferometric technique that can experimentally capture the Husimi Q-function directly.  In addition, this method can be adapted beyond synchronization to other studies that rely on coherence measurements, even weak ones. By measuring the NMR system in the process of reaching the steady state,  we investigate  transient as well as  steady state behaviour of the system under the influence of the external drive. 
We also explore the robustness of synchronization against detuning frequency and drive strength via the Arnold tongue. 

This article is organized as follows: in Sec. \ref{Theory}, we introduce the spin-system and outline the theory as well as measures of quantum synchronization. In Sec. \ref{Methodology}, we explain the NMR architecture, the experimental setup, and the numerical model. 
The interferometric measurement of Husimi distribution (IMHD) is described in Sec. \ref{NMRsyncexp}. Following this, we discuss the corresponding experimental results in Sec. \ref{sec:results}.  Finally, we conclude in Sec. \ref{Conclusions}.

\begin{figure}
	\includegraphics[trim=1.2cm 8cm 1cm 2cm,width=9cm,clip=]{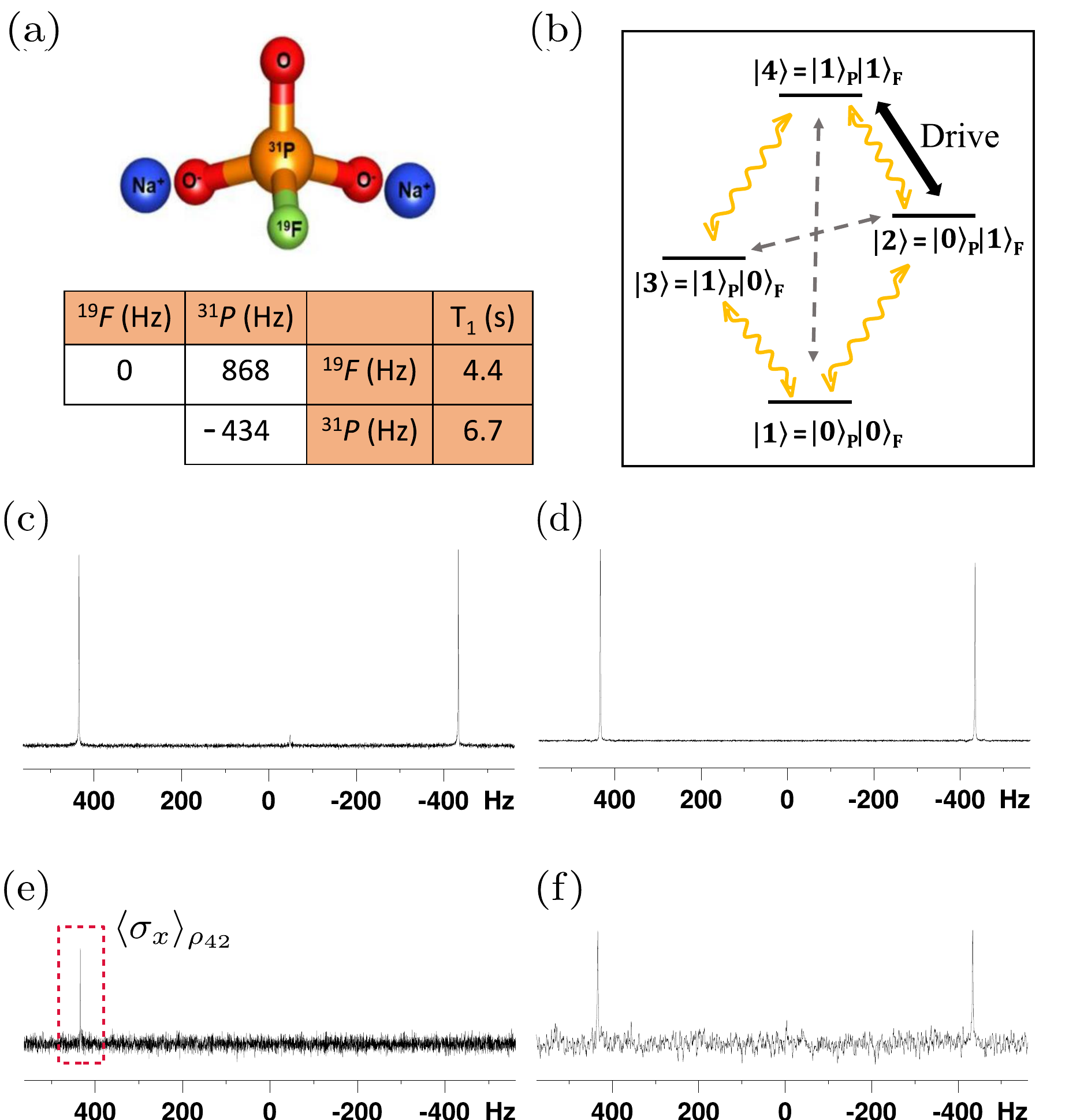}
	\caption{(a) Sodium fluorophosphate molecular structure with its Hamiltonian parameters and the spin-lattice relaxation time constants (T$_1$) shown in the table. The diagonal elements represent the offset, while off-diagonal element represents the scalar $J$ coupling constant. Figure (b) represents the energy levels of the two-qubit system with four non-degenerate energy eigenstates.  The numerical model presented here considers only the single-quantum relaxation pathways (yellow curved arrows) and ignores the zero- and double-quantum pathways (dashed grey arrows). Figures (c,d) show reference NMR spectra of $^{31}$P and $^{19}$F spins respectively, each obtained with a $90^\circ$ pulse on thermal equilibrium state, while (e) shows the $^{31}$P NMR signal after 100 s drive on the $\ket{2} \leftrightarrow \ket{4}$ transition. Intensity of the $\rho_{42}$ element is indicated in the dashed box, which is used for the Arnold tongue analysis. Figure (f) shows the $^{19}$F NMR spectrum at the end of interferometric circuit from which we can directly extract the Husimi distribution at a particular $\theta$ and $\phi$ values.}
	\label{molecule}
\end{figure}

\section{Synchronization of a four-level system}
\label{Theory}
We consider a non-degenerate four-level system composed of $^{19}$F and $^{31}$P nuclei in sodium fluorophospate molecule, as shown in Fig. \ref{molecule}(a,b) and study the phase synchronization of the system with an external drive. The extent of phase localization of the system in a state $\rho$ can be quantified using the Husimi-Kano Q representation function \cite{husimi1940some,kano1965new}. For the given four-level system it can be defined as 
\begin{equation}
Q(\theta_1,\theta_2,\theta_3,\phi_1,\phi_2,\phi_3) =
 \frac{24}{\pi^3}\dmel{\hat{n}_4}{\rho},
\label{Qfn}
\end{equation}
where 
$$\ket{\hat{n}_4}= (C_{\theta_1}, e^{i\phi_1}S_{\theta_1}C_{\theta_2}, e^{i\phi_2}S_{\theta_1}S_{\theta_2}C_{\theta_3}, e^{i\phi_3}S_{\theta_1}S_{\theta_2}S_{\theta_3})^T,$$
is $SU(4)$ coherent state \cite{perelomov1986,nemoto2000generalized,mathur2002n} with $C_{\theta_i}=\cos(\theta_i/2)$ and $S_{\theta_i}=\sin(\theta_i/2)$. A brief review of spin coherent states is presented in Appendix \ref{appendix1}. The normalization of the Q-function arises from the completeness relation of coherent state $\ket{\hat{n}_4}$ defined as
\begin{align}
\int \proj{\hat{n}_4} d\mu = \frac{\pi^3}{24} \mathbbm{1},
\end{align}
where the integration is taken over the Haar measure  and volume element given by
\cite{nemoto2000generalized} 
$$d\mu =d\theta_1 d\theta_2 d\theta_3 d\phi_1 d\phi_2 d\phi_3C_{2\theta_1}S^5_{2\theta_1}C_{2\theta_2}S_{2\theta_2}^3C_{2\theta_3}S_{2\theta_3}.$$ 

For a system with internal Hamiltonian $ H_0 = \sum_{i=1}^{4}\omega_i \ket{i}\bra{i}$ having characteristic frequencies $\omega_i$, the spin coherent state evolves as
\begin{align}
e^{-iH_0t}\ket{\hat{n}_4}
\rightarrow
\left(
\begin{array}{c}
C_{\theta_1}
\\
e^{i(\phi_1-\omega_{21}t)}S_{\theta_1} C_{\theta_2}
\\
e^{i(\phi_2-\omega_{31}t)}S_{\theta_1}S_{\theta_2}C_{\theta_3}
\\
e^{i(\phi_3-\omega_{41}t)}S_{\theta_1}S_{\theta_2}S_{\theta_3}
\end{array}
\right),
\end{align}
where $\omega_{i1} = \omega_{i}-\omega_1$.
Under the free evolution of internal Hamiltonian, $\phi_i$s represent the free phases oscillating with the respective frequencies $\omega_{i1}$ while $\theta_i$s govern the populations of the system which remains fixed.  
In the absence of any external perturbation, the steady state reached by the thermalization has uniform phase distribution, and the Husimi Q-function remains  independent of $(\phi_1,\phi_2,\phi_3 )$. This points to the fact that the given system has free phases available. Along with the existence of free phases, a system needs to be nonlinear in nature to exhibit limit cycle behaviour. Since the dynamics of a multilevel quantum system is generally nonlinear, the availability of free phases makes it a perfect candidate to study synchronization. 
On application of a weak external drive the system may develop a definite phase relationship with the drive, thus resulting in a localized phase distribution. This phenomenon, known as phase synchronization, is captured and quantified by
\begin{align}
S(\phi_1,\phi_2,\phi_3) &= \int
d\Theta~ Q(\theta_1,\theta_2,\theta_3,\phi_1,\phi_2,\phi_3)  - \frac{1}{(2\pi)^3},
\end{align}
where $d\Theta=d\theta_1 d\theta_2 d\theta_3 C_{2\theta_1}S^5_{2\theta_1}C_{2\theta_2}S_{2\theta_2}^3C_{2\theta_3}S_{2\theta_3}$. For the four-level system considered here, the above expression leads to
\begin{align}
S(\phi_1,\phi_2,\phi_3) &=  \frac{1}{16\pi^2} \mathrm{Re}  \left[\rho_{43}e^{i\phi_1}+\rho_{42}e^{i\phi_2}
+ \rho_{41}e^{i\phi_3} \right.
\nonumber \\
 &\left.+ \rho_{32}e^{i(\phi_2-\phi_1)} 
+ \rho_{31}e^{i(\phi_3-\phi_1)}  + \rho_{21}e^{i(\phi_3-\phi_2)}\right],
\label{seqn}
\end{align}
where $\rho_{ij} = \expec{i|\rho|j}$.  
Under the effect of thermal baths and in the absence of external perturbations/drive, the system reaches its thermal equilibrium state, that is diagonal in the energy eigenbasis of the internal Hamiltonian. The Husimi Q-function for a diagonal state reduces to a uniform distribution showing no phase localization. Therefore the thermal state constitutes a valid limit cycle state which is stable to external perturbation and has free phases \cite{jaseem2020quantum,jaseem2020generalized}. Furthermore, the synchronization measure $S(\phi_1,\phi_2,\phi_3)$ registers zero for such limit cycle state.
\\
\\
\textit{System with drive - } Let us now study the entrainment of the given limit cycle oscillator with an external drive.  A drive having strength $\Omega$ and frequency $\omega_d$ is applied on the
$\ket{2} \leftrightarrow \ket{4}$ transition (see Fig. \ref{molecule} b). The Hamiltonian describing the system with the drive is given by
\begin{equation}
H = \sum_{i=1}^{4} \omega_i \ket{i}\bra{i} + \Omega
\left(
\ket{2}\bra{4}e^{i\omega_d t}+\ket{4}\bra{2}e^{-i\omega_d t}
\right).
\label{Hlabframe}
\end{equation}
In the rotating frame of the drive described by the unitary transformation $$\exp({i\{(\omega_d + \omega_2) \proj{4} + \omega_3  \proj{3} + \omega_2\proj{2} + \omega_1 \proj{1}\}t}),$$ the total Hamiltonian becomes
$H_R = \Delta \proj{4} + \Omega(\outpr{2}{4} + \outpr{4}{2})$, where $\Delta=(\omega_4 - \omega_2) - \omega_d$. 
For the $\ket{2}\leftrightarrow \ket{4}$ drive considered here, the only surviving coherence under steady state is $\rho_{42}$ (and its adjoint $\rho_{24}$). 
Therefore dropping all other coherence terms, and relabelling $\phi_2$ as $\phi$, the Q-function defined in Eq.~(\ref{Qfn}) can be expressed as
\begin{align}
\label{Husimi}
Q(\theta_1,\theta_2,\theta_3,\phi) &=  \mathrm{Re}\left(\rho_{42}e^{i\phi}\right)S_{\theta_1} S_{\theta_2}C_{\theta_3}+ \rho_{11}S^2_{\theta_1} S^2_{\theta_2}S^2_{\theta_3}\nonumber\\
&+ \rho_{22}S^2_{\theta_1} S^2_{\theta_2} C^2_{\theta_3}+\rho_{33}S^2_{\theta_1} C^2_{\theta_2}+\rho_{44}C^2_{\theta_1} .
\end{align}
Since the drive is not sensitively affecting the levels $\ket{1}$ and $\ket{3}$, without any loss of generality we can effectively set  $\theta_2 = \pi$, $\theta_3 = 0$ and replace $\theta_1 \rightarrow \theta$ in the above equation, which further simplifies the Husimi function to 
\begin{align}
Q(\theta,\phi) 
= \rho_{44} C^2_{\theta}
+ \mathrm{Re}\left(\rho_{42}e^{i\phi}\right) S_{2\theta} + \rho_{22} S^2_{\theta}.
\label{eq:Q_final}
\end{align}
From the above equation it is clear that non-zero $\rho_{42}$ leads to the localization of the corresponding phase variable $\phi$ in Husimi Q-function. Such a state corresponds to the synchronized state.
The synchronization measure $S(\phi_1,\phi_2,\phi_3)$  corresponding to Eq.~(\ref{seqn}) also reduces to 
\begin{equation}
\label{Smaxeq}
S(\phi) = \frac{\mathrm{Re}\left(\rho_{42}e^{i\phi}\right)}{16\pi^2}~\mbox{and}~
\max\left[S(\phi)\right] = \frac{\left|\rho_{42}\right|}{16\pi^2},
\end{equation}
which is non-zero only for $\rho_{42}\neq 0$, similar to \cite{jaseem2020quantum}.

\section{NMR Methodology}
\label{Methodology}
\subsection{The spin system and the drive}
In this work, we consider a two-qubit system formed by the spin-1/2 nuclei $^{19}$F and $^{31}$P of sodium fluorophosphate molecule dissolved in D$_2$O solvent (5.3 mg of solute in 600 $\mu$l of solvent) and maintained at an ambient temperature of 298 K. 
All experiments were performed on a high resolution Bruker 500 MHz NMR spectrometer operating at a magnetic field strength of $B_0 = 11.4$ T.  Fig. \ref{molecule}(a) displays the spin system and its Hamiltonian parameters including scalar spin-spin coupling constant ($J$-coupling), as well as the RF offsets $\nu_P$, $\nu_F$ with respect to the Larmor frequencies $\omega_P = -\gamma_P B_0$,
$\omega_F = -\gamma_F B_0$, where $\gamma_i$ are the gyromagnetic ratios.  Note that $\nu_P$ is set to $-J/2$, while $\nu_F$ is set to zero.
Fig. \ref{molecule}(b) shows the Zeeman energy level diagram of this two-qubit system. The lab-frame Hamiltonian is of the form
\begin{align}
H_\mathrm{NMR} = \omega_P I_z^P + \omega_F I_z^F + 2 \pi J I_z^P I_z^F.
\end{align}
In NMR systems, at ambient temperatures, the thermal energy is much greater than the Zeeman energy splitting. Hence, an \textit{n}-qubit system at thermal equilibrium is in a highly mixed state given by 
\begin{eqnarray}
\rho^\mathrm{eq} = \exp\left(\frac{-H_0}{k_BT}\right) = \mathbbm{1}/2^n + \sum_{i}\epsilon_i I_z^i,
\end{eqnarray}
where $\epsilon = \hbar \gamma_i B_0/(2^nk_BT) \sim 10^{-5}$ is the purity factor. The population distribution in the high-temperature limit follows the Boltzmann distribution. The inherent relaxation (\textit{T$_1$}) mechanism facilitates dissipation, and helps establish equilibrium population distribution, which forms a stable limit cycle with free phases as discussed in the previous section. The thermal equilibrium spectra of $^{31}$P and $^{19}$F spins are shown in Fig. \ref{molecule} (c,d) respectively.

To realize synchronization, we apply a weak drive of amplitude $\Omega \approx 0.1$ Hz (explained in further detail in Sec. \ref{sec:optimumdrive}) selectively on the $\ket{2} \leftrightarrow \ket{4}$ transition as indicated in Fig. \ref{molecule} (b). 
The total Hamiltonian in the doubly rotating frame is given by
\begin{align}
H_\mathrm{tot} &= H_0 + V ~\mbox{where,} 
\nonumber \\
H_0 &= -2\pi \nu_P I_z^P  + 2 \pi J I_z^P I_z^F~\mbox{and}~ V= 2 \pi \Omega I_y^P.
\end{align}
The first term with $\nu_P = -J/2$ makes the drive on-resonant with the $\ket{2} \leftrightarrow \ket{4}$ transition of the $^{31}$P qubit 
that prepares $\rho_{42}$ coherence (see Fig. \ref{molecule}(e)), while the dissipation mechanism redistributes the populations of the system.

\subsection{Steady state dynamics \label{sec:model}}

We now describe a simple model to estimate the instantaneous state of the two-qubit system under different drive conditions. The time evolution of a quantum system is given by the master equation \cite{breuer2002theory}
\begin{equation}
\frac{d\rho}{dt} = -i[H_0 + V,\rho] + \mathcal{D}[\hat{O}_+]\rho + \mathcal{D}[\hat{O}_-]\rho
\label{mastereqn}
\end{equation}
where $H_0$ is the internal Hamiltonian of the system and $V$ represents the external drive. The thermal bath at a temperature $T$ is modeled by the Lindblad superoperators $\mathcal{D}[\hat{O}_+]\rho + \mathcal{D}[\hat{O}_-]\rho$. The single
quantum upward  transitions (indicated by yellow curves in Fig. \ref{molecule}(b)) are described by the jump operator $\mathcal{D}[\hat{O}+]\rho$ accompanied by their corresponding transition probabilities and rates. The equivalent downward transitions are described by the Hermitian conjugate $\mathcal{D}[\hat{O}_-]\rho$. The transition probabilities $p_{ij}^{kl}$ can be identified with upward or downward transitions of a  particular spin.  We estimate these probabilities via the Fermionic bath model \cite{ghosh2012fermionic}.  Relabelling the upward \& downward transitions of $n$th spin as $p_{n\pm}$,  we may write
\begin{align}
p_{n+} = \frac{1}{e^{4\epsilon_n}+1}~~
\mbox{and} ~~ p_{n-} = 1- p_{n+},
\end{align}
which help maintain detailed balance. The transition rate of each spin is related to the bath temperature $T$ via its spin-lattice relaxation time $T_1^i$ as $g_i = 2\pi/T_1^i$ (see Fig. \ref{molecule}(a) for $T_1$ values). The explicit forms of the Lindblad superoperators are given in Appendix \ref{appendix2}.

 The master equation in Eq.~(\ref{mastereqn}) is linear in $\rho$ and can be written as follows
\begin{equation}
\lket{\Dot{\rho}} = 
 \mathcal{L} \lket{\rho} = (\mathcal{L}_0+\mathcal{L}_V)\lket{\rho}
\label{modelmastereqn}
\end{equation}
where $\mathcal{L}$ represents the Liouvillian superoperator describing the open system dynamics and $\lket{\rho}$ is the vectorised form of density matrix in Liouville space. The mathematical form of $\mathcal{L}$ can be obtained by applying the transformation 
$B\rho C \rightarrow C^* \otimes B \lket{\rho}$ to the master equation where $\lket{\rho}$ is obtained by vertically stacking the columns of density matrix \cite{suri2018speeding,manzano2020short,gyamfi2020fundamentals,solanki2021role}.

The Liouvillian superoperator $\mathcal{L}$ can be decomposed into two parts as shown in Eq.~(\ref{modelmastereqn}). The first part $\mathcal{L}_0$ defines the dynamics of system in absence of external drive (V=0) and is given by
\begin{eqnarray}
\mathcal{L}_0 &=& -i(I\otimes H_0-H_0^*\otimes I)  +\sum_{j=+,-}O_j^T\otimes O_j\nonumber \\ &-&\frac{1}{2}(I\otimes O_j^\dagger O_j+O_j^T O_j^* \otimes I).
\end{eqnarray}
The second term $\mathcal{L}_V$ represents the superoperator corresponding to an external perturbation $V$ which is given by
\begin{equation}
    \mathcal{L}_V=-i(I\otimes V-V^*\otimes I).
\end{equation}


The instantaneous state $\rho(t)$ after solving Eq.~(\ref{modelmastereqn}) is given by
\begin{align}
\rho(t) = e^{\mathcal{L}t} \rho_0,
\end{align}
where $\rho_0$ represents the initial state of the system. The steady state is given by converging solution of Eq.~(\ref{modelmastereqn}), defined as $\rho_{ss}=\lim_{t\rightarrow \infty}e^{\mathcal{L}t} \rho_0$. Therefore the steady state corresponds to the eigenstate of Liouvillian superoperator having zero eigenvalue \cite{navarrete2015open}.
In the absence of external drive, one obtains a diagonal steady state,  $\rho^{eq}=\mathrm{diag}\{\rho_{44}^{eq},\rho_{33}^{eq},\rho_{22}^{eq},\rho_{11}^{eq}\}$, with  elements following the thermal distribution.
In the presence of an external drive $V$, the steady state corresponds to the eigenstate of $\mathcal{L}$ having zero eigenvalue  and is of the form
\begin{eqnarray}
\rho^\mathrm{ss} = 
\left[
\begin{array}{cccc}
\rho^\mathrm{ss}_{44} & 0 & \rho^\mathrm{ss}_{42} & 0 \\
0 & \rho^\mathrm{ss}_{33} & 0 & 0 \\
\rho^\mathrm{ss}_{24} & 0 & \rho^\mathrm{ss}_{22} & 0 \\
0 & 0 & 0 &  \rho^\mathrm{ss}_{11}
\end{array}
\right].
\end{eqnarray}
Here, the basis states are ordered according to decreasing energy eigenvalues.  Also note that since the external drive is very weak, the steady state attained is close to the limit cycle, \textit{i.e.}, $\rho^\mathrm{ss}_{ii} \approx \rho^\mathrm{eq}_{ii}$. 

We would like to emphasise that it is a highly non-trivial task to measure all the Lindblad dissipation operators for the NMR system \cite{boulant2003robust}. The dynamics in the system is vastly more complex than the model considered here can capture. The set of operators comprising this model is very minimal and intended only to capture the effective dissipation effects. We note that a full description of the NMR system involves many more terms not considered in our minimal model.

\subsection{Optimum drive amplitude}
\label{sec:optimumdrive}
\begin{figure}[t]
    \centering
    \includegraphics[trim=5cm 6.5cm 5cm 7cm,width=9cm,clip=]{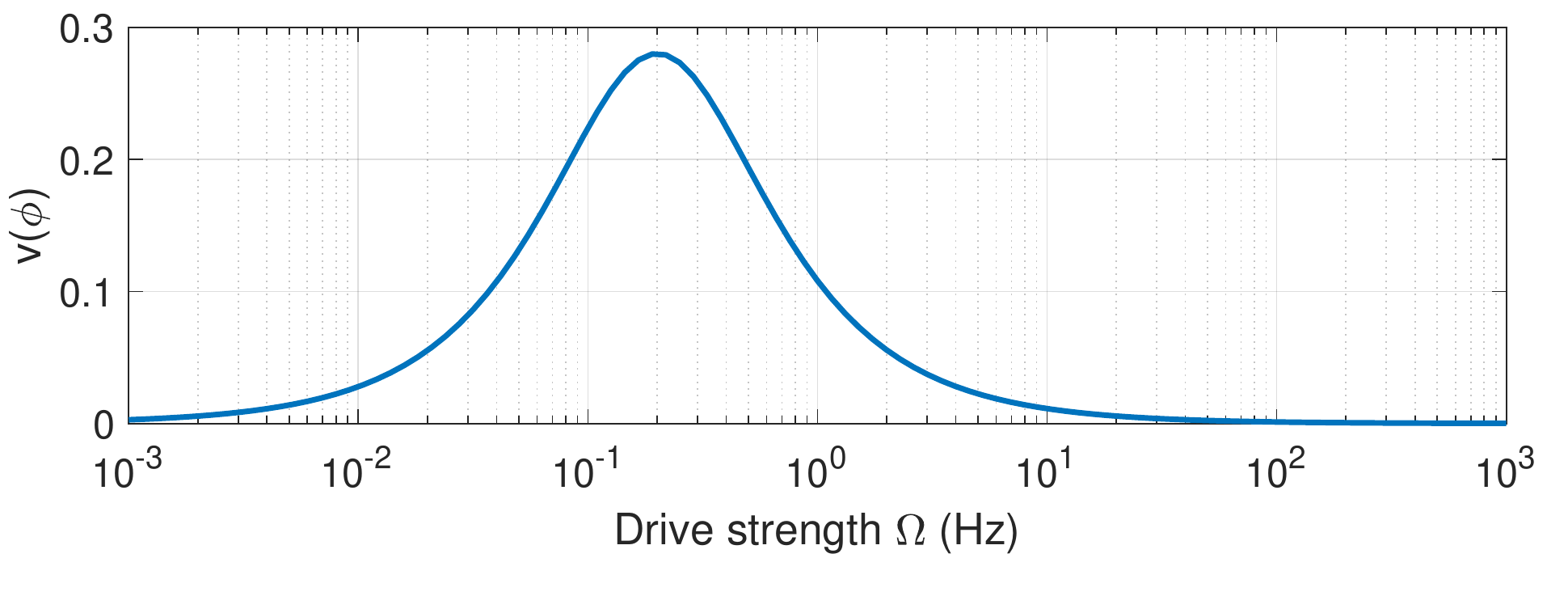}
    \caption{The dependence of  visibility of phase-localization versus drive amplitude $\Omega$, showing an optimum strength at around 0.1 Hz.}
    \label{optds}
\end{figure}

The strength of the drive is crucial to realising synchronization in any system. A very weak drive will barely perturb the system, while a very strong drive will induce forced behaviour and alter the limit cycle. It is hence vital to limit the drive strength within an appropriate regime to achieve synchronization. For our system, we estimate the optimum drive strength using the numerical model described in Sec. \ref{sec:model}. We vary the drive strength and numerically obtain the corresponding steady-state, for which we estimate the visibility of the Husimi distribution 
\begin{align}
\mathrm{v}(\phi) = \frac{\mathrm{max}(Q_\phi)-\mathrm{min}(Q_\phi)}{\mathrm{max}(Q_\phi)+\mathrm{min}(Q_\phi)},
\end{align}
where $Q_\phi = \sum_{\theta} Q(\theta,\phi)$ 
Note that visibility quantifies the extent of phase-localization.
The result is shown in Fig. \ref{optds}. The drive amplitude was varied from $10^{-3}$ Hz to $10^3$ Hz. We can see that the visibility function is negligible for very low powers and then peaks at around $10^{-1}$ Hz before dropping to zero at higher drive amplitudes. We thus chose a drive strength of 0.1 Hz for our subsequent Husimi distribution experiments. The experimental method to calibrate the low amplitude drive pulse is explained in the Appendix \ref{appendix3}.

\section{Interferometric measurement of Husimi distribution (IMHD)}
\label{NMRsyncexp}
Theoretically, the instantaneous state can be directly predicted by solving the master equation, following which the Husimi function can be evaluated using Eq.~(\ref{Qfn}). Experimentally, the standard way to extract Husimi distribution is by using Eq.~(\ref{Qfn}) after carrying out QST to estimate the steady state density matrix $\rho$ \cite{krithika2019nmr}. However, there are two unrelated issues that plague tomographic measurements in quantum systems.
One relates to the fact that the experimental complexity of QST scales exponentially with the system size \cite{chuang1998bulk,shukla2013ancilla}, each component of which has to be measured repeatedly. This introduces deleterious noise sources when multiple experiments are needed to perform QST. Having said that, for small quantum systems such as a four-level system, full tomography is routine. This leads us to the second issue, which is that even in small system sizes considered here, QST of steady states turned out to be highly inefficient.  The reason being QST relies on determining each element of the density matrix via a linear combination of expectation values of a set of observables.  If the expectation values vary over  large magnitudes, the dynamic range problem prevents the estimation of faintest elements.
The IMHD method described here avoids the dynamic range problem and directly extracts the Husimi function at each $(\theta,\phi)$ value in a single experiment without requiring the elaborate QST protocols.  This allowed us to efficiently observe and quantify synchronization even after a very long drive as explained below.

\begin{figure}[b]
	\flushleft(a) 
	\hspace*{0.7cm}
	\Qcircuit @C=1em @R=1em {
    & & ~\mbox{Thermal bath} & & & & \\
	\lstick{^{19}\mathrm{F}}  \barrier[0.1em]{1}  & \qw & \ctrl{1} & \barrier[-2.1em]{1} \qw & \gate{H} & \ctrl{1} & \meterB{\sigma_x}
		\\
	\lstick{^{31}\mathrm{P}} & \qw & \gate{ \ket{2}\leftrightarrow\ket{4} ~\mathrm{drive}} & \qw & \gate{U^\dagger_{\theta,\phi}} & \gate{Z} & \qw \gategroup{1}{3}{3}{3}{0.6em}{.} \gategroup{2}{3}{3}{3}{1em}{_\}} \gategroup{2}{5}{3}{7}{0.8em}{_\}} \\
	\rho(0)=\rho^\mathrm{eq} & & ~~~~~~~~~~~~~t ~~~~~~~~~~~~~\rho(t) & & & \tau << t& 
	} \\ ~ \\
	\flushleft(b)
	\includegraphics[trim=1.5cm 7cm 1cm 3cm,width=8.5cm,clip=]{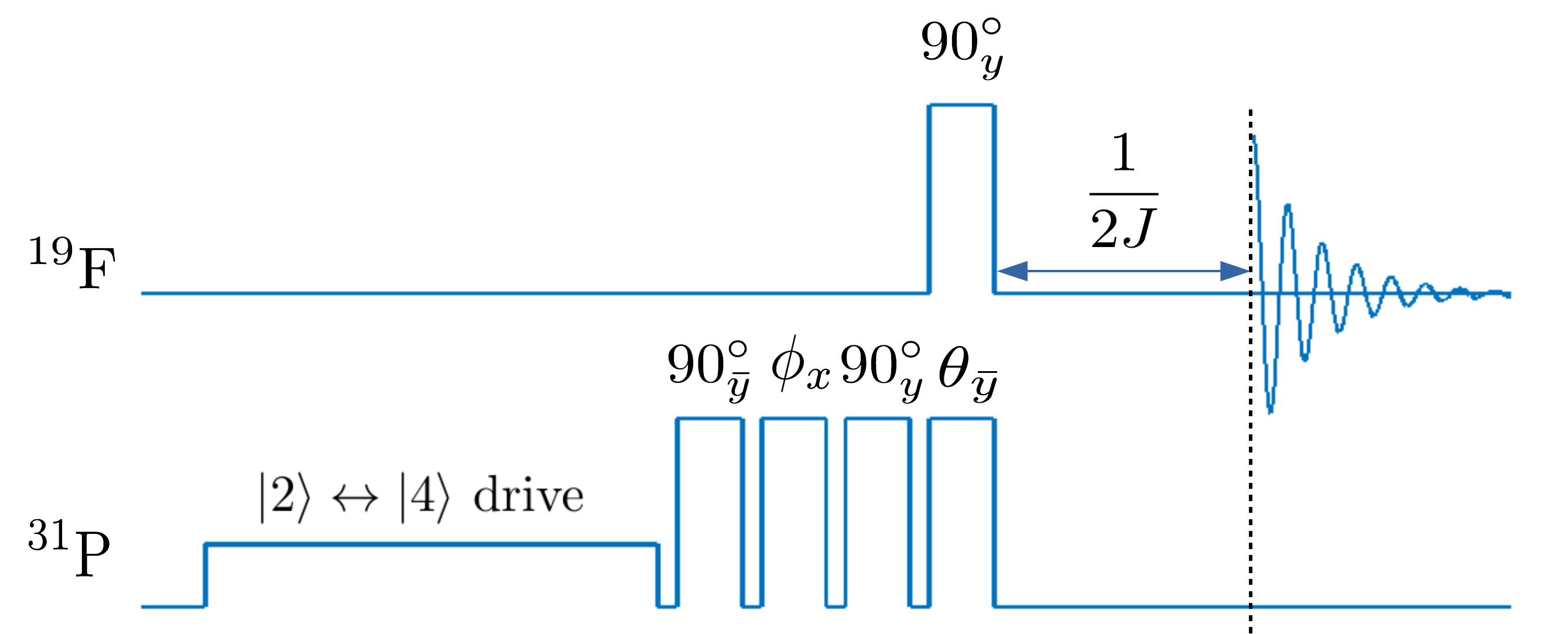}
	\caption{(a) IMHD circuit for direct measurement of Husimi distribution where the drive duration $t$ is much longer than the measurement sequence time $\tau$ and
	(b) the corresponding NMR pulse sequence.}
	\label{circuit}
\end{figure}

The 
circuit diagram representing the experiment to read the Husimi Q-function values  is shown in Fig. \ref{circuit}(a). Here, $^{19}$F qubit acts as the ancillary system, which along with $^{31}$P qubit begin from a thermal equilibrium state $\rho^\mathrm{eq}$. 
The long weak-drive responsible for synchronization is applied on the $\ket{2} \leftrightarrow \ket{4}$ transition of the $^{31}$P qubit. 
As part of the interferometer, we now apply a Hadamard operator on $^{19}$F qubit and prepare its superposition. Meanwhile, we apply the gate $U^\dagger_{\theta,\phi}$ on $^{31}$P qubit. Subsequently, we apply a controlled phase-gate $\mathbbm{1}_P \otimes \ket{0}\bra{0}_F + \sigma_z^P \otimes \ket{1}\bra{1}_F$.

Starting from the extremal state $\ket{0,0} = \ket{\theta = 0,\phi=0}$, we can generate the spin coherent state
\begin{align}
\ket{\theta,\phi}  = U_{\theta,\phi}  \ket{0,0}= \left[e^{-i\phi S_z} e^{-i\theta S_y}\otimes \mathbbm{1}^F\right] \ket{0,0}.
\end{align}
The corresponding IMHD reading is
\begin{align}
Q_\mathrm{IF}(\theta,\phi) &= \dmel{\theta,\phi}{\rho^\mathrm{ss}} 
= \mathrm{Tr}\left(\rho^\mathrm{ss} \proj{\theta,\phi}\right) \nonumber \\
&= \mathrm{Tr}\left(\rho^\mathrm{ss} U_{\theta,\phi}\proj{0,0}U_{\theta,\phi}^\dagger\right) \nonumber \\
&= \mathrm{Tr}\left(U_{\theta,\phi}^\dagger \rho^\mathrm{ss} U_{\theta,\phi}\proj{0,0}\right) \nonumber \\
&= \expv{\proj{0,0}}{\rho_{\theta,\phi}},
\end{align}
where $\rho_{\theta,\phi} = U_{\theta,\phi}^\dagger \rho^\mathrm{ss} U_{\theta,\phi}$. 
\begin{figure*}[htp]
	\centering
	\flushleft(a) \\
	\includegraphics[trim=3.5cm 10.4cm 2.5cm 6.0cm,width=19cm,clip=]{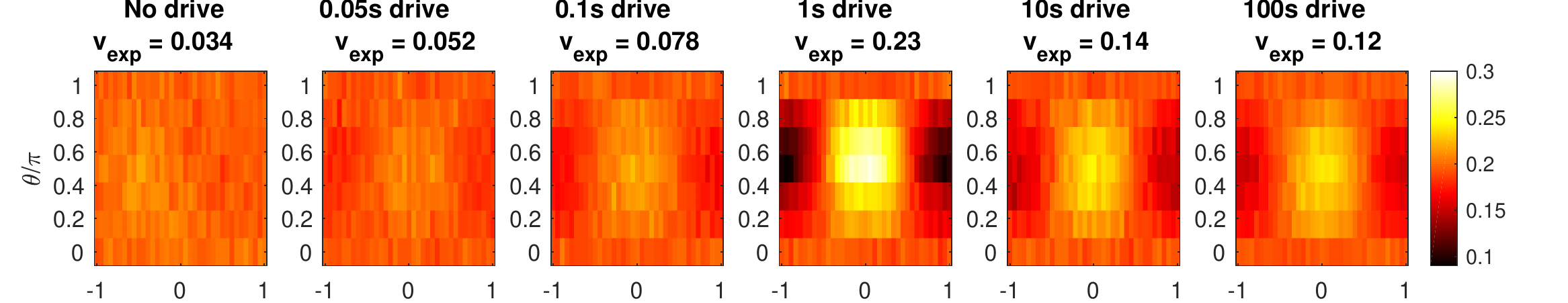}
	\flushleft(b) \\
	\includegraphics[trim=3.5cm 5.3cm 2.5cm 10.8cm,width=19cm,clip=]{husimi_all_5.pdf}
	\caption{The full phase space Husimi distribution values for different drive durations to capture transient and steady state behaviour in the subspace of interest obtained (a) by experiments  and (b) by the numerical model. The values mentioned in the titles indicate the visibility factors of the Husimi distributions. We can see that in the absence of drive, there is no localization in the phase space. Upon applying the drive, in the transient regime (upto 10 s), the phase begins to localize gradually, showing the strongest localization experimentally at a drive duration of 1 s and saturating after 10 s.}
	\label{results}
\end{figure*}
 
The NMR pulse sequence for IMHD is shown in Fig. \ref{circuit}(b).
After the  $\ket{2}\leftrightarrow \ket{4}$ drive, we implement a pseudo-Hadamard operator using a $90^\circ_y$ pulse on the $^{19}$F spin.
While $U_{\theta}^\dagger = e^{i\theta I_y}$ can be realized by a single $\theta_{\bar{y}}$ pulse,  $U_{\phi}^\dagger = e^{i\phi I_z} = e^{-i \frac{\pi}{2} I_y} e^{-i\phi I_x} e^{i \frac{\pi}{2} I_y}$ is realized by a sequence of three pulses.
The measurement of $\proj{0,0}$ projector can be realized via a controlled phase-gate, which can be implemented, up to a phase-factor, by a simple free evolution under the system Hamiltonian $2\pi J I_z^P I_z^F$ for a time duration of $1/2J$. A subsequent measurement of transverse magnetization of the $^{19}$F spin yields the NMR signal
\begin{align}
s = \mathrm{Re}\expv{\sigma_x^F}{} =& \frac{1}{2}\left\{C_{2\theta} \left[\rho_{11}^\mathrm{ss} -\rho_{22}^\mathrm{ss} - \rho_{33}^\mathrm{ss} + \rho_{44}^\mathrm{ss}\right]\right. \nonumber \\
&+ \left.S_{2\theta} ~ \mathrm{Re}(e^{-i\phi}\rho_{24}^\mathrm{ss}+e^{i\phi}\rho_{42}^\mathrm{ss})\right\}.
\label{signal}
\end{align}
One can verify from Eq.~(\ref{Husimi}) and Eq.~(\ref{signal}) that
\begin{align}
Q_\mathrm{IF}(\theta,\phi) &= \frac{24}{\pi^3}\left[\frac{1 + 2s}{2} - \left(\rho_{11}^\mathrm{ss}C^2_{\theta} + \rho_{33}^\mathrm{ss} S^2_{\theta}\right)\right] \nonumber \\
&=Q(\theta,\phi).
\label{signal2husimi}
\end{align}
Here $\rho_{11}^\mathrm{ss} \simeq \rho_{11}^\mathrm{eq}$ and $\rho_{33}^\mathrm{ss} \simeq \rho_{33}^\mathrm{eq}$ are the populations of undriven levels and can be estimated from the thermal equilibrium population distribution. For the highly mixed NMR systems, $\rho_{11} \simeq \rho_{33} \simeq 1/4$, so that,
\begin{align}
Q_\mathrm{IF}(\theta,\phi) &= 
\frac{24}{\pi^3}\left[\frac{1 + 2s}{2} - \frac{1}{4}\right] = \frac{24}{\pi^3}\left[s + \frac{1}{4}\right].
\label{simpleQif}
\end{align}
 
 
Thus, the $^{19}$F signal after the IMHD circuit of Fig. \ref{circuit} can directly measure the Husimi distribution values. One such spectrum for a particular drive duration and $(\theta,\phi)$ values is shown in Fig. \ref{molecule}(f).
Note that one may use a third spin as an ancilla qubit for the interferometer circuit, but it will open up additional dissipation pathways.  To 
minimise such decoherence channels, we have used one of the system spins itself as an ancilla qubit.


\section{Results}
\label{sec:results}
The experimental IMHD results at thermal equilibrium as well as at various drive durations are shown in Fig. \ref{results}.  The signal measured at the end of the IMHD circuit (Fig. \ref{circuit}) is directly used to estimate the Husimi distributions using Eq.~(\ref{simpleQif}). The title of each sub-figure indicates the value of the corresponding visibility factor of the Husimi distribution.


\paragraph*{Limit cycle:} 
In the absence of drive, the system is in thermal equilibrium with the environment. Husimi distribution of the thermal equilibrium state $\rho^\mathrm{eq}$ is shown in Fig. \ref{results} (a), which is uniform throughout the phase space with no phase localization, and accordingly vanishingly small visibility factor. Therefore $\rho^\mathrm{eq}$ is a valid limit cycle with free phase, as is expected.

\paragraph*{Onset of synchronization:}
To study synchronization in the system, we apply a weak transition-selective  drive of strength $\Omega = 0.1$ Hz on-resonant with the $\ket{2}\leftrightarrow \ket{4}$  transition as shown in Fig. \ref{molecule}(b). The equilibration time of the quantum system is approximately 10 s, which means that dynamics captured under 10 s can be considered transient whereas for timescales much larger than 10 s we observe steady states of the system. The drive is applied for different durations which enabled us to investigate the transient and steady state dynamics in the subspace of interest.
The experimentally measured Husimi distributions for various drive durations are shown in Fig.~\ref{results} (a). The corresponding distributions for the instantaneous states predicted by the numerical model (described in section \ref{sec:model}) are shown in Fig. \ref{results} (b). 
We can see that the system gradually develops phase localization with the drive before reaching a steady state. 
For short drive durations, i.e., for 50 ms, 100 ms we observe a weak phase localization, which captures the transient dynamics. In this transient regime, the experimental phase distribution reached a peak localization strength at 1 s drive duration.
The system then reaches a steady state with the phase localization stabilizing for drive durations roughly above 10 s.  It remains synchronized even up till 100 s, which is more than ten times the drive period as well as ten times the $T_1$ relaxation time constants of the system spins. 

The experimental phase localization pattern and visibility factors match fairly well with those predicted by the  numerical model.  While
an elaborate relaxation model, accounting for all the Lindblad dissipation operators and other experimental imperfections such as RF inhomogeneity might explain the observed deviations, the current minimal model nonetheless captures the essential signatures of synchronization.

\paragraph*{Arnold tongue:}

\begin{figure}
	\includegraphics[trim=2cm 4.5cm 2cm 3cm,width=9cm,clip=]{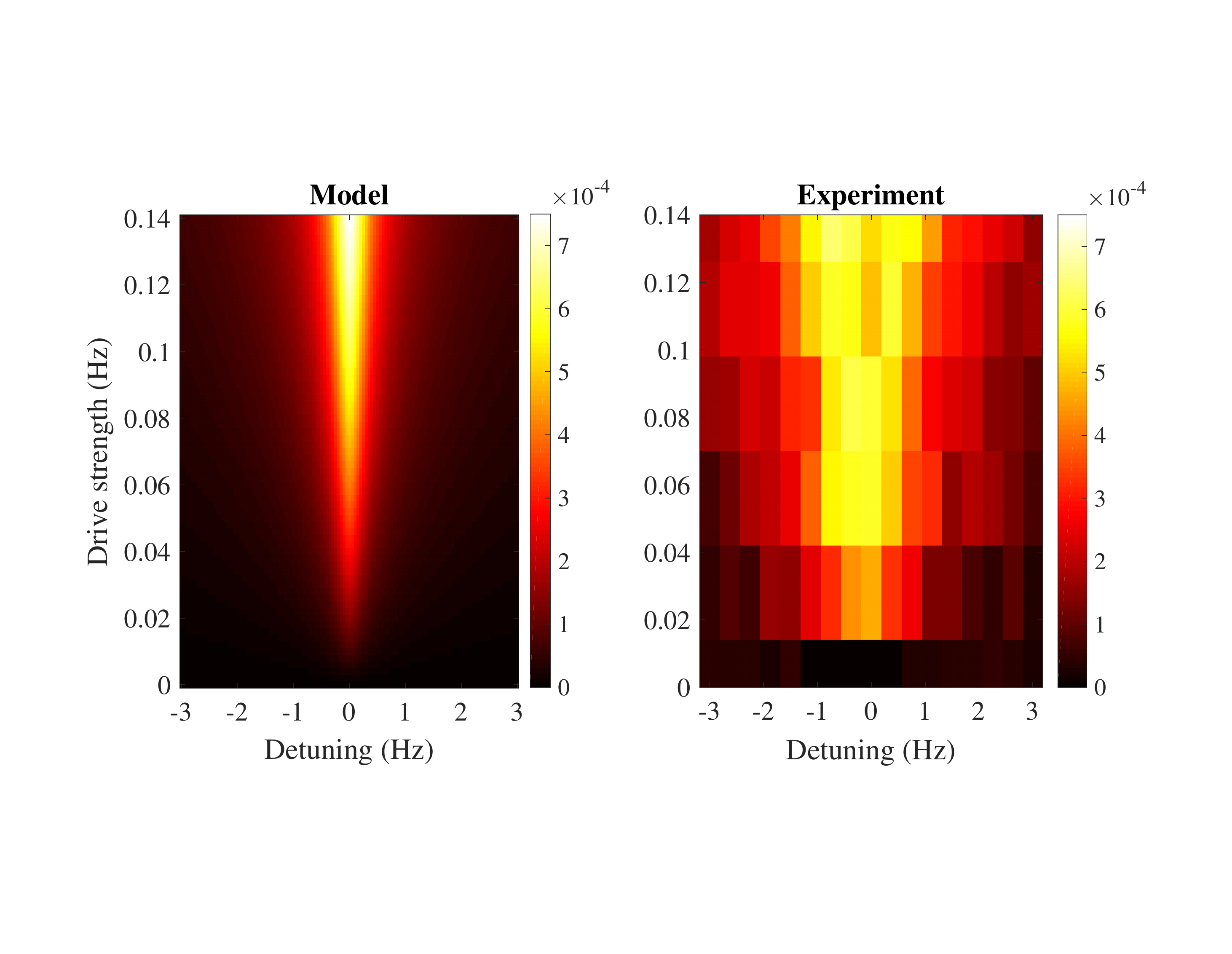}
	\caption{Arnold tongue behaviour of the system. The system shows strongest synchronization for on-resonant drive and strong drive strengths, while off-resonant drives are unable to synchronize with the transition of interest even for large driving strengths.}
	\label{arnold}
\end{figure}
 Arnold tongue is a quintessential test of synchronization that probes robustness of phase-localization against small changes in drive strengths and drive detuning \cite{pikovsky2003synchronization}. As the drive strength is increased, the region of synchronization can be shown to be wider, leading to the familiar ``tongue" shape.
 Here, we vary the drive strength of the 100 s weak drive, and its resonance-offset by 3 Hz about the resonance value. The Arnold tongue is quantified using the maximum of the synchronization measure, $\max\left[S(\phi)\right]$,  and in the present case it is only dependent on the $\rho_{42}$ element of the density matrix, as evident from Eq.~(\ref{Smaxeq}). Experimentally, this can be directly measured from the intensity of $^{31}$P NMR spectrum (see Fig. \ref{molecule}(e)). 
Fig. \ref{arnold} compares the result of the Arnold tongue experiments with that of the numerical prediction (from the model described in section \ref{sec:model}).
For strong drive strength and near resonance conditions, the degree of synchronization is higher since the drive has more efficient perturbative effect on the system. On the other hand, for larger detuning and weaker drives, the extent of synchronization is weaker. 
Despite an overall correspondence between the experimental and the predicted profiles, we see a higher spread along the detuning axis in the experimental data, which can be attributed to the limitations of the minimal numerical model and experimental imperfections.


\section{Summary and Outlook}
\label{Conclusions}
 Recently there has been an increasing interest in studying the synchronization of quantum systems with external drives under suitable conditions.  In this work, we experimentally demonstrated synchronization in a two-qubit system using NMR architecture. We use the Husimi distribution function as a witness for synchronization. We applied a weak transition selective drive on one of the spins and observed the gradual onset of phase-localization via Husimi distribution.  
 In the absence of any external drive, the system evolves under its internal Hamiltonian as well as inherent relaxation mechanisms, thereby settling to its thermal equilibrium state.  The corresponding Husimi distribution pattern has no phase-preference, and hence the thermal equilibrium state forms the valid limit cycle of the system. 
 
 An interesting issue that arises with the study of steady state dynamics of open quantum systems relates to the fact that NMR signals are proportional to the population differences. Since the populations are in a pseudo-spin state, their effective difference dips below the noise threshold, rendering steady state population tomography difficult in such systems until now. Here we reported an interferometric technique to directly extract the Husimi distribution values by reading the signal of the undriven spin.  The interferometric method is significantly more efficient compared to the standard method based on quantum state tomography. 
 
 To establish the robustness of synchronization, we investigated the response of the system to changes in drive strengths and drive detuning.  The resulting phase-localization measure of the system exhibited the expected Arnold tongue behaviour. We compared the experimental results with a minimalistic $T_1$ relaxation model of the two-qubit nuclear spin system, which could capture the general features of the experimental data. 
 
 This work demonstrates the suitability of NMR architecture for quantum synchronization studies and also opens up avenues for further exploration of the phenomenon in larger spin systems as well as under a variety of interactions and drive conditions. One can envisage NMR systems being useful for studying the implications of synchronization in areas such as spectroscopy, quantum computing and quantum thermodynamics.

\section*{Acknowledgements}
VRK acknowledges valuable discussions with Conan Alexander. SV acknowledges support from a Government of India DST-SERB Early Career Research Award (ECR/2018/000957) and DST-QUEST grant number DST/ICPS/QuST/Theme-4/2019.
TSM acknowledges funding from
DST/ICPS/QuST/2019/Q67.

\appendix
\section{Spin coherent states}
\label{appendix1}
Spin coherent states are routinely used to study synchronization since their evolutions are closest to  classical trajectories \cite{lee2015visualizing}.  Furthermore, coherent states exhibit stable oscillations as explained below.
For a spin-1/2 particle, the  $z$ projection of the spin angular momentum $S_z$ is quantized into 2s+1 levels $\ket{S,m_s}$ with eigenvalues $m_s = -1/2,1/2$.
The spin coherent state for this spin in SU(2) group can be described as a rotation of the extremal state $\ket{S,S}$ such that \cite{radcliffe1971some,klauder1985generalized,nemoto2000generalized}
\begin{align}
\ket{\hat{n}_2} = \ket{\theta,\phi} = \mathrm{e}^{-i\phi S_z} \mathrm{e}^{-i\theta S_y} \ket{S,S} 
= 
\left(
\begin{array}{c}
C_{\theta_{1}}
\\
e^{i\phi} S_{\theta_{1}}
\end{array}
\right),
\end{align}
where $\theta \in [0,\pi]$, $\phi \in [0,2\pi]$, $C_{\theta} = \cos(\theta/2)$, and $S_{\theta} = \sin(\theta/2)$.
It is obvious that spin-1/2 coherent states can be mapped to points on the surface of the Bloch sphere. The recursive construction of spin coherent state vectors for an $n$-level non-degenerate system is  \cite{nemoto2000generalized,perelomov1986}
\begin{align}
\ket{\hat{n}_n} =
\left(
\begin{array}{c}
 C_{\theta}
\\
0
\\
\vdots
\\
0
\end{array}
\right) + 
e^{i\phi} S_{\theta}
\left(
\begin{array}{c}
0
\\
\\
\ket{\hat{n}_{n-1}}
\\
\\
\end{array}
\right).
\end{align}
Note that each lower-level vector $\ket{\hat{n}_{n-1}}$ is made up of a new pair of angular variables.
From the above, we obtain the $SU(4)$ coherent states in the form
\begin{align}
\ket{\hat{n}_4} =
\left(
\begin{array}{c}
C_{\theta_{1}}
\\
e^{i\phi_1} S_{\theta_{1}} C_{\theta_{2}}
\\
e^{i\phi_2}S_{\theta_{1}}S_{\theta_{2}}C_{\theta_{3}}
\\
e^{i\phi_3}S_{\theta_{1}}S_{\theta_{2}}S_{\theta_{3}}
\end{array}
\right).
\label{coherentstate}
\end{align}
From the above definition, we can see that populations are governed by the parameters $(\theta_1,\theta_2,\theta_3)$, and phases by the parameters $(\phi_1,\phi_2,\phi_3)$.

\section{Lindblad relaxation operators}
\label{appendix2}
The explicit form of the single-quantum jump operators considered for the four-level system shown in Fig. \ref{molecule}(b) are explained here. The single-quantum transition of the first qubit implies that the system in a state $\ket{i}\ket{j}$ goes to the state $\ket{k}\ket{l}$, where $k = i\pm1$, and $l = j$. The same can be extended to the single-quantum transition of the second qubit. The Lindblad superoperator for upward transitions of the combined two-qubit system is hence given by $\mathcal{D}[\hat{O}_+]\rho=\sum_{(k-i = 1,l=j)} D[\mathcal{O}_{ij}^{kl}]\rho + \sum_{(k=i,l-j = 1)} D[\mathcal{O}_{ij}^{kl}]\rho$ where $D[\hat{\mathcal{O}}]\rho = \hat{\mathcal{O}}\rho\hat{\mathcal{O}}^\dagger - \{\hat{\mathcal{O}}^\dagger\hat{\mathcal{O}},\rho\}/2$ and jump operator $\mathcal{O}_{ij}^{kl}$ is defined as $\mathcal{O}_{ij}^{kl}=\sqrt{g_{ij}^{kl}p_{ij}^{kl}}~ \ket{k}\ket{l}\bra{i}\bra{j}$. Similarly, we can write $\mathcal{D}[\hat{O}_-]\rho=\sum_{(k-i=1,l=j)} D[\mathcal{O}^{ij}_{kl}]\rho + \sum_{(k=i,l-j = 1)} D[\mathcal{O}^{ij}_{kl}]\rho$, where $\mathcal{O}^{ij}_{kl}=(\mathcal{O}_{ij}^{kl})^\dagger$.  The coefficients of the jump operators corresponding to each qubit determine the transition rate given by $g_{ij}^{kl}$, and the transition probability given by $p_{ij}^{kl}$. The transition rate is dependent on the bath temperature via spin-lattice relaxation time constant $T_1$, and the transition probabilities are derived from a fermionic bath model, as explained in the main text.

\section{Very low power RF calibration}
\label{appendix3}

\begin{figure}[h!]
    \vspace{0.5cm}
	\flushleft(a)\\
	\centering
	\includegraphics[trim=10cm 8cm 7cm 2.5cm,width=6cm,clip=]{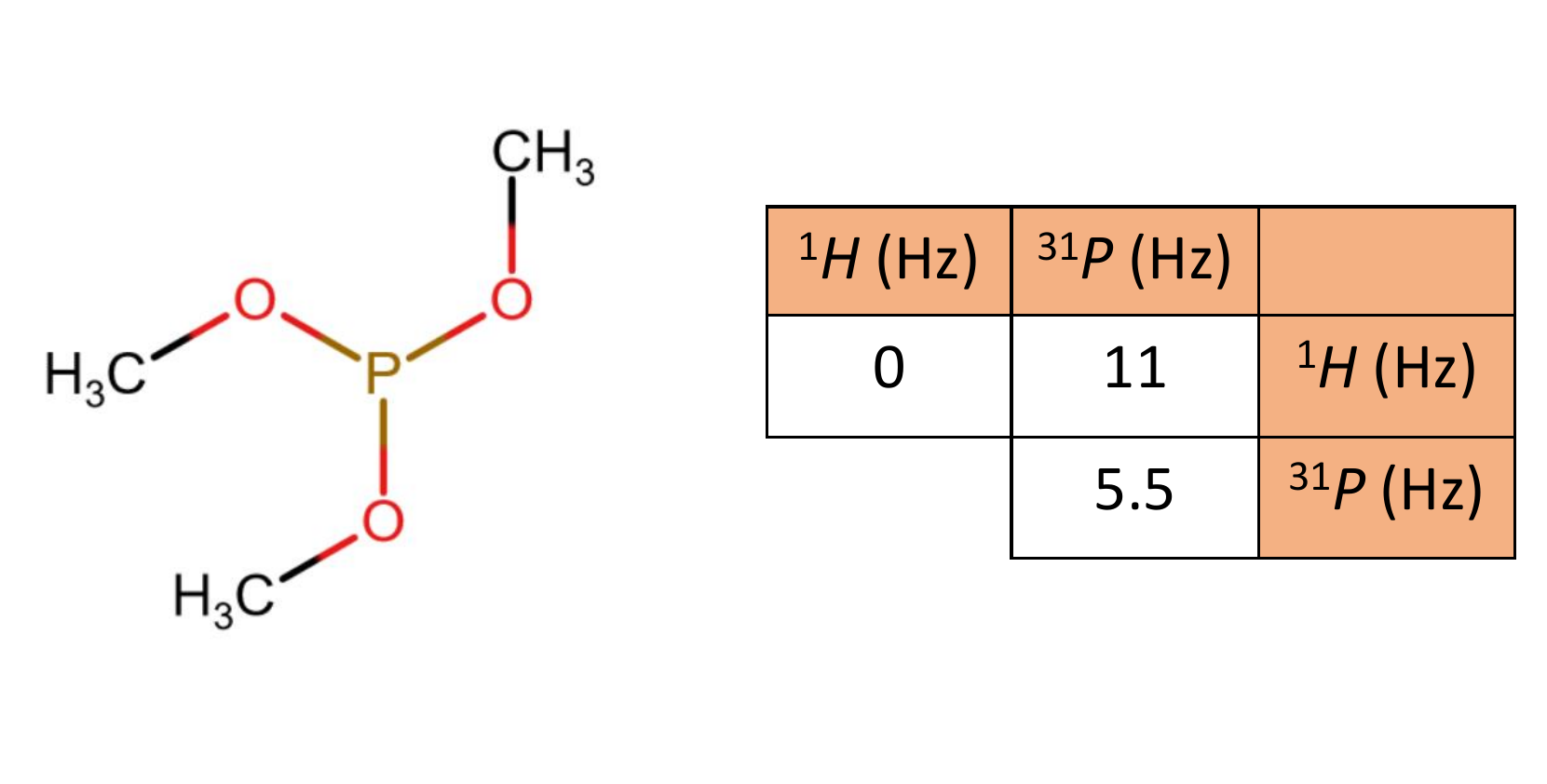}
	\caption{TMP molecule and Hamiltonian parameters.}
	\label{tmp}
 	
	\flushleft(b)\\
	\includegraphics[trim=2cm 2cm 2cm 2cm,width=8cm,clip=]{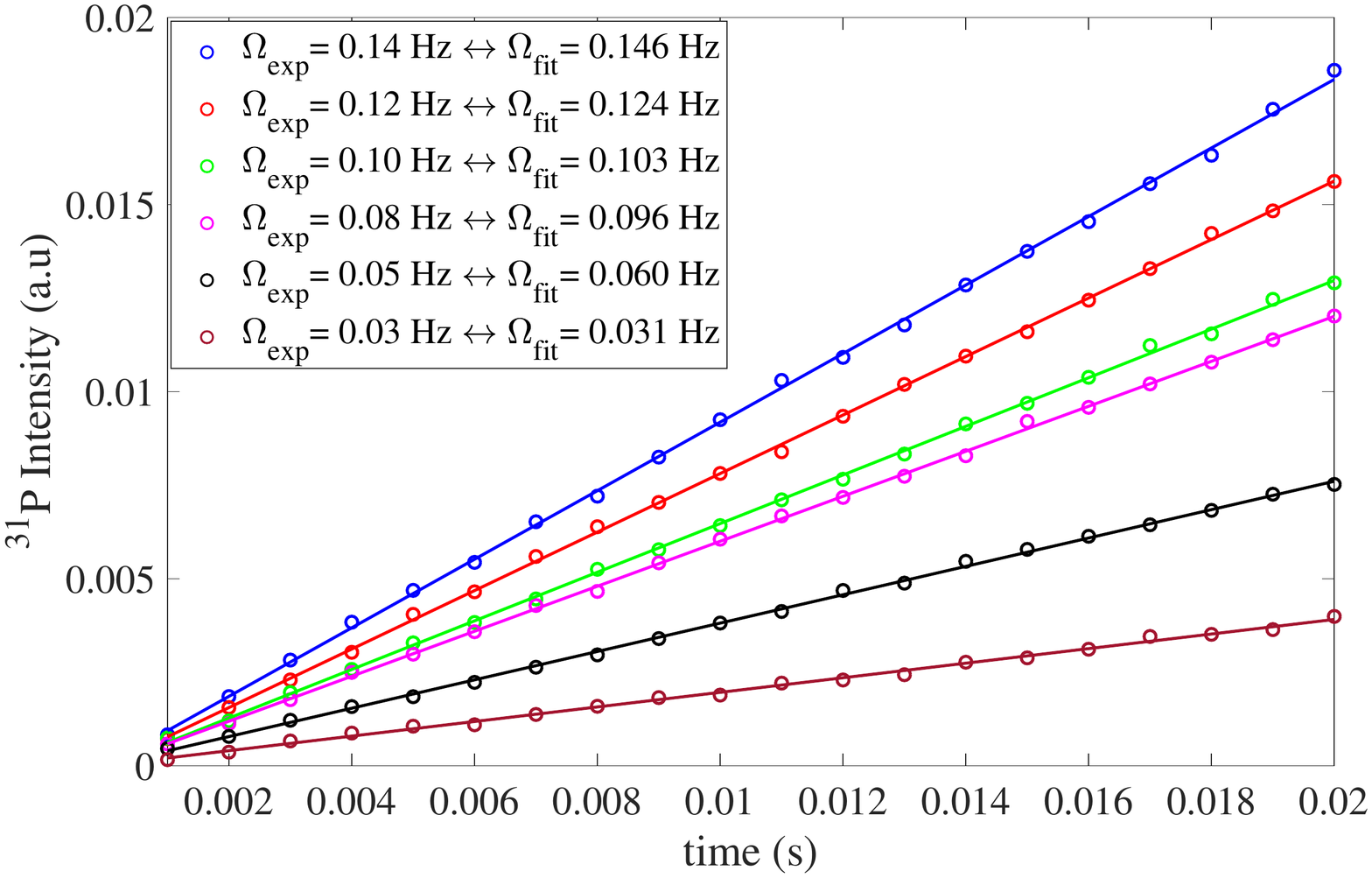}
	\caption{Low power calibration using trimethylphosphite. The intensity of $^{31}$P spin after decoupling protons is shown as a function of time for various low amplitude pulses. We can see that the response is linear in this regime. The expected and exactly obtained values from back-calculation of the linear fit are shown in the legend.}
	\label{lowpowercalib}
\end{figure}

As discussed in section \ref{sec:optimumdrive}, observation of phase-synchronization requires a drive of amplitude $10^{-1}$ Hz.  The calibration of such very weak RF is nontrivial, since the signal to noise ratio is negligibly small.
For this purpose, we used another spin-system trimethylphosphite dissolved in dimethylsulphoxide solvent. We specifically chose this sample because of its star-topology with $^{31}$P at the centre coupled to nine identical $^1$H nuclei with a scalar J coupling constant of 11 Hz, as shown in Fig. \ref{tmp}. Such geometry is highly preferable since the magnetization of the nine high $\gamma$ proton nuclei can be easily transferred to the central $^{31}$P nucleus via INEPT and further algorithmic cooling, thus boosting its signal \cite{VaradPRA}. This gives better signal to noise ratio and hence more accurate calibration values. 
For very low amplitudes, the signal $\propto \sin(\Omega t)\approxeq \Omega t$, and therefore we expect a linear dependence.
The calibration results are shown in Fig. \ref{lowpowercalib}. As we can see, the intensity of the $^{31}$P spectrum after decoupling protons varies linearly with time for the low amplitude pulses. Thus by fitting a linear function to the resulting curve, we can back-calculate the exact amplitude of the pulse. The estimated and exact values from back calculation of the linear fit are shown in the legend. These low power pulses were used in the Husimi distribution estimation and Arnold tongue experiments.
\bibliography{references}

\begin{thebibliography}{50}%
\makeatletter
\providecommand \@ifxundefined [1]{%
 \@ifx{#1\undefined}
}%
\providecommand \@ifnum [1]{%
 \ifnum #1\expandafter \@firstoftwo
 \else \expandafter \@secondoftwo
 \fi
}%
\providecommand \@ifx [1]{%
 \ifx #1\expandafter \@firstoftwo
 \else \expandafter \@secondoftwo
 \fi
}%
\providecommand \natexlab [1]{#1}%
\providecommand \enquote  [1]{``#1''}%
\providecommand \bibnamefont  [1]{#1}%
\providecommand \bibfnamefont [1]{#1}%
\providecommand \citenamefont [1]{#1}%
\providecommand \href@noop [0]{\@secondoftwo}%
\providecommand \href [0]{\begingroup \@sanitize@url \@href}%
\providecommand \@href[1]{\@@startlink{#1}\@@href}%
\providecommand \@@href[1]{\endgroup#1\@@endlink}%
\providecommand \@sanitize@url [0]{\catcode `\\12\catcode `\$12\catcode
  `\&12\catcode `\#12\catcode `\^12\catcode `\_12\catcode `\%12\relax}%
\providecommand \@@startlink[1]{}%
\providecommand \@@endlink[0]{}%
\providecommand \url  [0]{\begingroup\@sanitize@url \@url }%
\providecommand \@url [1]{\endgroup\@href {#1}{\urlprefix }}%
\providecommand \urlprefix  [0]{URL }%
\providecommand \Eprint [0]{\href }%
\providecommand \doibase [0]{https://doi.org/}%
\providecommand \selectlanguage [0]{\@gobble}%
\providecommand \bibinfo  [0]{\@secondoftwo}%
\providecommand \bibfield  [0]{\@secondoftwo}%
\providecommand \translation [1]{[#1]}%
\providecommand \BibitemOpen [0]{}%
\providecommand \bibitemStop [0]{}%
\providecommand \bibitemNoStop [0]{.\EOS\space}%
\providecommand \EOS [0]{\spacefactor3000\relax}%
\providecommand \BibitemShut  [1]{\csname bibitem#1\endcsname}%
\let\auto@bib@innerbib\@empty
\bibitem [{\citenamefont {Pikovsky}\ \emph {et~al.}(2003)\citenamefont
  {Pikovsky}, \citenamefont {Kurths}, \citenamefont {Rosenblum},\ and\
  \citenamefont {Kurths}}]{pikovsky2003synchronization}%
  \BibitemOpen
  \bibfield  {author} {\bibinfo {author} {\bibfnamefont {A.}~\bibnamefont
  {Pikovsky}}, \bibinfo {author} {\bibfnamefont {J.}~\bibnamefont {Kurths}},
  \bibinfo {author} {\bibfnamefont {M.}~\bibnamefont {Rosenblum}},\ and\
  \bibinfo {author} {\bibfnamefont {J.}~\bibnamefont {Kurths}},\ }\href@noop {}
  {\emph {\bibinfo {title} {Synchronization: a universal concept in nonlinear
  sciences}}},\ \bibinfo {number} {12}\ (\bibinfo  {publisher} {Cambridge
  university press},\ \bibinfo {year} {2003})\BibitemShut {NoStop}%
\bibitem [{\citenamefont {Lee}\ and\ \citenamefont
  {Sadeghpour}(2013)}]{lee2013quantum}%
  \BibitemOpen
  \bibfield  {author} {\bibinfo {author} {\bibfnamefont {T.~E.}\ \bibnamefont
  {Lee}}\ and\ \bibinfo {author} {\bibfnamefont {H.~R.}\ \bibnamefont
  {Sadeghpour}},\ }\bibfield  {title} {\bibinfo {title} {Quantum
  synchronization of quantum van der pol oscillators with trapped ions},\
  }\href {https://doi.org/10.1103/PhysRevLett.111.234101} {\bibfield  {journal}
  {\bibinfo  {journal} {Physical review letters}\ }\textbf {\bibinfo {volume}
  {111}},\ \bibinfo {pages} {234101} (\bibinfo {year} {2013})}\BibitemShut
  {NoStop}%
\bibitem [{\citenamefont {Hush}\ \emph {et~al.}(2015)\citenamefont {Hush},
  \citenamefont {Li}, \citenamefont {Genway}, \citenamefont {Lesanovsky},\ and\
  \citenamefont {Armour}}]{hush2015spin}%
  \BibitemOpen
  \bibfield  {author} {\bibinfo {author} {\bibfnamefont {M.~R.}\ \bibnamefont
  {Hush}}, \bibinfo {author} {\bibfnamefont {W.}~\bibnamefont {Li}}, \bibinfo
  {author} {\bibfnamefont {S.}~\bibnamefont {Genway}}, \bibinfo {author}
  {\bibfnamefont {I.}~\bibnamefont {Lesanovsky}},\ and\ \bibinfo {author}
  {\bibfnamefont {A.~D.}\ \bibnamefont {Armour}},\ }\bibfield  {title}
  {\bibinfo {title} {Spin correlations as a probe of quantum synchronization in
  trapped-ion phonon lasers},\ }\href
  {https://doi.org/10.1103/PhysRevA.91.061401} {\bibfield  {journal} {\bibinfo
  {journal} {Physical Review A}\ }\textbf {\bibinfo {volume} {91}},\ \bibinfo
  {pages} {061401} (\bibinfo {year} {2015})}\BibitemShut {NoStop}%
\bibitem [{\citenamefont {Nigg}(2018)}]{nigg2018observing}%
  \BibitemOpen
  \bibfield  {author} {\bibinfo {author} {\bibfnamefont {S.~E.}\ \bibnamefont
  {Nigg}},\ }\bibfield  {title} {\bibinfo {title} {Observing quantum
  synchronization blockade in circuit quantum electrodynamics},\ }\href
  {https://doi.org/10.1103/PhysRevA.97.013811} {\bibfield  {journal} {\bibinfo
  {journal} {Physical Review A}\ }\textbf {\bibinfo {volume} {97}},\ \bibinfo
  {pages} {013811} (\bibinfo {year} {2018})}\BibitemShut {NoStop}%
\bibitem [{\citenamefont {Xu}\ \emph {et~al.}(2014)\citenamefont {Xu},
  \citenamefont {Tieri}, \citenamefont {Fine}, \citenamefont {Thompson},\ and\
  \citenamefont {Holland}}]{xu2014synchronization}%
  \BibitemOpen
  \bibfield  {author} {\bibinfo {author} {\bibfnamefont {M.}~\bibnamefont
  {Xu}}, \bibinfo {author} {\bibfnamefont {D.~A.}\ \bibnamefont {Tieri}},
  \bibinfo {author} {\bibfnamefont {E.~C.}\ \bibnamefont {Fine}}, \bibinfo
  {author} {\bibfnamefont {J.~K.}\ \bibnamefont {Thompson}},\ and\ \bibinfo
  {author} {\bibfnamefont {M.~J.}\ \bibnamefont {Holland}},\ }\bibfield
  {title} {\bibinfo {title} {Synchronization of two ensembles of atoms},\
  }\href {https://doi.org/10.1103/PhysRevLett.113.154101} {\bibfield  {journal}
  {\bibinfo  {journal} {Physical review letters}\ }\textbf {\bibinfo {volume}
  {113}},\ \bibinfo {pages} {154101} (\bibinfo {year} {2014})}\BibitemShut
  {NoStop}%
\bibitem [{\citenamefont {Ludwig}\ and\ \citenamefont
  {Marquardt}(2013)}]{ludwig2013prl}%
  \BibitemOpen
  \bibfield  {author} {\bibinfo {author} {\bibfnamefont {M.}~\bibnamefont
  {Ludwig}}\ and\ \bibinfo {author} {\bibfnamefont {F.}~\bibnamefont
  {Marquardt}},\ }\bibfield  {title} {\bibinfo {title} {Quantum many-body
  dynamics in optomechanical arrays},\ }\href
  {https://doi.org/10.1103/PhysRevLett.111.073603} {\bibfield  {journal}
  {\bibinfo  {journal} {Phys. Rev. Lett.}\ }\textbf {\bibinfo {volume} {111}},\
  \bibinfo {pages} {073603} (\bibinfo {year} {2013})}\BibitemShut {NoStop}%
\bibitem [{\citenamefont {Walter}\ \emph {et~al.}(2014)\citenamefont {Walter},
  \citenamefont {Nunnenkamp},\ and\ \citenamefont
  {Bruder}}]{walter2014quantum}%
  \BibitemOpen
  \bibfield  {author} {\bibinfo {author} {\bibfnamefont {S.}~\bibnamefont
  {Walter}}, \bibinfo {author} {\bibfnamefont {A.}~\bibnamefont {Nunnenkamp}},\
  and\ \bibinfo {author} {\bibfnamefont {C.}~\bibnamefont {Bruder}},\
  }\bibfield  {title} {\bibinfo {title} {Quantum synchronization of a driven
  self-sustained oscillator},\ }\href
  {https://doi.org/10.1103/PhysRevLett.112.094102} {\bibfield  {journal}
  {\bibinfo  {journal} {Physical review letters}\ }\textbf {\bibinfo {volume}
  {112}},\ \bibinfo {pages} {094102} (\bibinfo {year} {2014})}\BibitemShut
  {NoStop}%
\bibitem [{\citenamefont {Walter}\ \emph {et~al.}(2015)\citenamefont {Walter},
  \citenamefont {Nunnenkamp},\ and\ \citenamefont
  {Bruder}}]{walter2015quantum}%
  \BibitemOpen
  \bibfield  {author} {\bibinfo {author} {\bibfnamefont {S.}~\bibnamefont
  {Walter}}, \bibinfo {author} {\bibfnamefont {A.}~\bibnamefont {Nunnenkamp}},\
  and\ \bibinfo {author} {\bibfnamefont {C.}~\bibnamefont {Bruder}},\
  }\bibfield  {title} {\bibinfo {title} {Quantum synchronization of two van der
  pol oscillators},\ }\href
  {https://doi.org/https://doi.org/10.1002/andp.201400144} {\bibfield
  {journal} {\bibinfo  {journal} {Annalen der Physik}\ }\textbf {\bibinfo
  {volume} {527}},\ \bibinfo {pages} {131} (\bibinfo {year}
  {2015})}\BibitemShut {NoStop}%
\bibitem [{\citenamefont {Li}\ \emph {et~al.}(2016)\citenamefont {Li},
  \citenamefont {Li},\ and\ \citenamefont {Song}}]{li2016quantum}%
  \BibitemOpen
  \bibfield  {author} {\bibinfo {author} {\bibfnamefont {W.}~\bibnamefont
  {Li}}, \bibinfo {author} {\bibfnamefont {C.}~\bibnamefont {Li}},\ and\
  \bibinfo {author} {\bibfnamefont {H.}~\bibnamefont {Song}},\ }\bibfield
  {title} {\bibinfo {title} {Quantum synchronization in an optomechanical
  system based on lyapunov control},\ }\href
  {https://doi.org/10.1103/PhysRevE.93.062221} {\bibfield  {journal} {\bibinfo
  {journal} {Physical Review E}\ }\textbf {\bibinfo {volume} {93}},\ \bibinfo
  {pages} {062221} (\bibinfo {year} {2016})}\BibitemShut {NoStop}%
\bibitem [{\citenamefont {Zhang}\ \emph {et~al.}(2012)\citenamefont {Zhang},
  \citenamefont {Wiederhecker}, \citenamefont {Manipatruni}, \citenamefont
  {Barnard}, \citenamefont {McEuen},\ and\ \citenamefont
  {Lipson}}]{zhang2012synchronization}%
  \BibitemOpen
  \bibfield  {author} {\bibinfo {author} {\bibfnamefont {M.}~\bibnamefont
  {Zhang}}, \bibinfo {author} {\bibfnamefont {G.~S.}\ \bibnamefont
  {Wiederhecker}}, \bibinfo {author} {\bibfnamefont {S.}~\bibnamefont
  {Manipatruni}}, \bibinfo {author} {\bibfnamefont {A.}~\bibnamefont
  {Barnard}}, \bibinfo {author} {\bibfnamefont {P.}~\bibnamefont {McEuen}},\
  and\ \bibinfo {author} {\bibfnamefont {M.}~\bibnamefont {Lipson}},\
  }\bibfield  {title} {\bibinfo {title} {Synchronization of micromechanical
  oscillators using light},\ }\href
  {https://doi.org/10.1103/PhysRevLett.109.233906} {\bibfield  {journal}
  {\bibinfo  {journal} {Physical review letters}\ }\textbf {\bibinfo {volume}
  {109}},\ \bibinfo {pages} {233906} (\bibinfo {year} {2012})}\BibitemShut
  {NoStop}%
\bibitem [{\citenamefont {Holmes}\ \emph {et~al.}(2012)\citenamefont {Holmes},
  \citenamefont {Meaney},\ and\ \citenamefont
  {Milburn}}]{holmes2012synchronization}%
  \BibitemOpen
  \bibfield  {author} {\bibinfo {author} {\bibfnamefont {C.~A.}\ \bibnamefont
  {Holmes}}, \bibinfo {author} {\bibfnamefont {C.~P.}\ \bibnamefont {Meaney}},\
  and\ \bibinfo {author} {\bibfnamefont {G.~J.}\ \bibnamefont {Milburn}},\
  }\bibfield  {title} {\bibinfo {title} {Synchronization of many nanomechanical
  resonators coupled via a common cavity field},\ }\href
  {https://doi.org/10.1103/PhysRevE.85.066203} {\bibfield  {journal} {\bibinfo
  {journal} {Physical Review E}\ }\textbf {\bibinfo {volume} {85}},\ \bibinfo
  {pages} {066203} (\bibinfo {year} {2012})}\BibitemShut {NoStop}%
\bibitem [{\citenamefont {Witthaut}\ \emph {et~al.}(2017)\citenamefont
  {Witthaut}, \citenamefont {Wimberger}, \citenamefont {Burioni},\ and\
  \citenamefont {Timme}}]{timme2017classical}%
  \BibitemOpen
  \bibfield  {author} {\bibinfo {author} {\bibfnamefont {D.}~\bibnamefont
  {Witthaut}}, \bibinfo {author} {\bibfnamefont {S.}~\bibnamefont {Wimberger}},
  \bibinfo {author} {\bibfnamefont {R.}~\bibnamefont {Burioni}},\ and\ \bibinfo
  {author} {\bibfnamefont {M.}~\bibnamefont {Timme}},\ }\bibfield  {title}
  {\bibinfo {title} {Classical synchronization indicates persistent
  entanglement in isolated quantum systems},\ }\href
  {https://doi.org/https://doi.org/10.1038/ncomms14829} {\bibfield  {journal}
  {\bibinfo  {journal} {Nature communications}\ }\textbf {\bibinfo {volume}
  {8}},\ \bibinfo {pages} {1} (\bibinfo {year} {2017})}\BibitemShut {NoStop}%
\bibitem [{\citenamefont {Roulet}\ and\ \citenamefont
  {Bruder}(2018{\natexlab{a}})}]{roulet2018quantum}%
  \BibitemOpen
  \bibfield  {author} {\bibinfo {author} {\bibfnamefont {A.}~\bibnamefont
  {Roulet}}\ and\ \bibinfo {author} {\bibfnamefont {C.}~\bibnamefont
  {Bruder}},\ }\bibfield  {title} {\bibinfo {title} {Quantum synchronization
  and entanglement generation},\ }\href
  {https://doi.org/10.1103/PhysRevLett.121.063601} {\bibfield  {journal}
  {\bibinfo  {journal} {Physical review letters}\ }\textbf {\bibinfo {volume}
  {121}},\ \bibinfo {pages} {063601} (\bibinfo {year}
  {2018}{\natexlab{a}})}\BibitemShut {NoStop}%
\bibitem [{\citenamefont {Jaseem}\ \emph
  {et~al.}(2020{\natexlab{a}})\citenamefont {Jaseem}, \citenamefont
  {Hajdu{\v{s}}ek}, \citenamefont {Vedral}, \citenamefont {Fazio},
  \citenamefont {Kwek},\ and\ \citenamefont
  {Vinjanampathy}}]{jaseem2020quantum}%
  \BibitemOpen
  \bibfield  {author} {\bibinfo {author} {\bibfnamefont {N.}~\bibnamefont
  {Jaseem}}, \bibinfo {author} {\bibfnamefont {M.}~\bibnamefont
  {Hajdu{\v{s}}ek}}, \bibinfo {author} {\bibfnamefont {V.}~\bibnamefont
  {Vedral}}, \bibinfo {author} {\bibfnamefont {R.}~\bibnamefont {Fazio}},
  \bibinfo {author} {\bibfnamefont {L.-C.}\ \bibnamefont {Kwek}},\ and\
  \bibinfo {author} {\bibfnamefont {S.}~\bibnamefont {Vinjanampathy}},\
  }\bibfield  {title} {\bibinfo {title} {Quantum synchronization in nanoscale
  heat engines},\ }\href {https://doi.org/10.1103/PhysRevE.101.020201}
  {\bibfield  {journal} {\bibinfo  {journal} {Physical Review E}\ }\textbf
  {\bibinfo {volume} {101}},\ \bibinfo {pages} {020201} (\bibinfo {year}
  {2020}{\natexlab{a}})}\BibitemShut {NoStop}%
\bibitem [{\citenamefont {Solanki}\ \emph {et~al.}(2021)\citenamefont
  {Solanki}, \citenamefont {Jaseem}, \citenamefont {Hajdu{\v{s}}ek},\ and\
  \citenamefont {Vinjanampathy}}]{solanki2021role}%
  \BibitemOpen
  \bibfield  {author} {\bibinfo {author} {\bibfnamefont {P.}~\bibnamefont
  {Solanki}}, \bibinfo {author} {\bibfnamefont {N.}~\bibnamefont {Jaseem}},
  \bibinfo {author} {\bibfnamefont {M.}~\bibnamefont {Hajdu{\v{s}}ek}},\ and\
  \bibinfo {author} {\bibfnamefont {S.}~\bibnamefont {Vinjanampathy}},\
  }\bibfield  {title} {\bibinfo {title} {Role of coherence and degeneracies in
  quantum synchronisation},\ }\href@noop {} {\bibfield  {journal} {\bibinfo
  {journal} {arXiv preprint arXiv:2104.04383}\ } (\bibinfo {year}
  {2021})}\BibitemShut {NoStop}%
\bibitem [{\citenamefont {Li}\ \emph {et~al.}(2017{\natexlab{a}})\citenamefont
  {Li}, \citenamefont {Li},\ and\ \citenamefont {Song}}]{li2017quantum}%
  \BibitemOpen
  \bibfield  {author} {\bibinfo {author} {\bibfnamefont {W.}~\bibnamefont
  {Li}}, \bibinfo {author} {\bibfnamefont {C.}~\bibnamefont {Li}},\ and\
  \bibinfo {author} {\bibfnamefont {H.}~\bibnamefont {Song}},\ }\bibfield
  {title} {\bibinfo {title} {Quantum synchronization and quantum state sharing
  in an irregular complex network},\ }\href
  {https://doi.org/10.1103/PhysRevE.95.022204} {\bibfield  {journal} {\bibinfo
  {journal} {Physical Review E}\ }\textbf {\bibinfo {volume} {95}},\ \bibinfo
  {pages} {022204} (\bibinfo {year} {2017}{\natexlab{a}})}\BibitemShut
  {NoStop}%
\bibitem [{\citenamefont {Hajdu{\v{s}}ek}\ \emph {et~al.}(2021)\citenamefont
  {Hajdu{\v{s}}ek}, \citenamefont {Solanki}, \citenamefont {Fazio},\ and\
  \citenamefont {Vinjanampathy}}]{hajduvsek2021seeding}%
  \BibitemOpen
  \bibfield  {author} {\bibinfo {author} {\bibfnamefont {M.}~\bibnamefont
  {Hajdu{\v{s}}ek}}, \bibinfo {author} {\bibfnamefont {P.}~\bibnamefont
  {Solanki}}, \bibinfo {author} {\bibfnamefont {R.}~\bibnamefont {Fazio}},\
  and\ \bibinfo {author} {\bibfnamefont {S.}~\bibnamefont {Vinjanampathy}},\
  }\bibfield  {title} {\bibinfo {title} {Seeding crystallization in time},\
  }\href@noop {} {\bibfield  {journal} {\bibinfo  {journal} {arXiv preprint
  arXiv:2111.04395}\ } (\bibinfo {year} {2021})}\BibitemShut {NoStop}%
\bibitem [{\citenamefont {Schleich}(2011)}]{schleich2011quantum}%
  \BibitemOpen
  \bibfield  {author} {\bibinfo {author} {\bibfnamefont {W.~P.}\ \bibnamefont
  {Schleich}},\ }\href {https://doi.org/10.1002/3527602976} {\emph {\bibinfo
  {title} {Quantum optics in phase space}}}\ (\bibinfo  {publisher} {John Wiley
  \& Sons},\ \bibinfo {year} {2011})\BibitemShut {NoStop}%
\bibitem [{\citenamefont {Galve}\ \emph {et~al.}(2017)\citenamefont {Galve},
  \citenamefont {Luca~Giorgi},\ and\ \citenamefont {Zambrini}}]{Galve2017}%
  \BibitemOpen
  \bibfield  {author} {\bibinfo {author} {\bibfnamefont {F.}~\bibnamefont
  {Galve}}, \bibinfo {author} {\bibfnamefont {G.}~\bibnamefont {Luca~Giorgi}},\
  and\ \bibinfo {author} {\bibfnamefont {R.}~\bibnamefont {Zambrini}},\
  }\bibinfo {title} {Quantum correlations and synchronization measures},\ in\
  \href {https://doi.org/10.1007/978-3-319-53412-1_18} {\emph {\bibinfo
  {booktitle} {Lectures on General Quantum Correlations and their
  Applications}}}\ (\bibinfo  {publisher} {Springer International Publishing},\
  \bibinfo {address} {Cham},\ \bibinfo {year} {2017})\ pp.\ \bibinfo {pages}
  {393--420}\BibitemShut {NoStop}%
\bibitem [{\citenamefont {Li}\ \emph {et~al.}(2017{\natexlab{b}})\citenamefont
  {Li}, \citenamefont {Zhang}, \citenamefont {Li},\ and\ \citenamefont
  {Song}}]{li2017properties}%
  \BibitemOpen
  \bibfield  {author} {\bibinfo {author} {\bibfnamefont {W.}~\bibnamefont
  {Li}}, \bibinfo {author} {\bibfnamefont {W.}~\bibnamefont {Zhang}}, \bibinfo
  {author} {\bibfnamefont {C.}~\bibnamefont {Li}},\ and\ \bibinfo {author}
  {\bibfnamefont {H.}~\bibnamefont {Song}},\ }\bibfield  {title} {\bibinfo
  {title} {Properties and relative measure for quantifying quantum
  synchronization},\ }\href {https://doi.org/10.1103/PhysRevE.96.012211}
  {\bibfield  {journal} {\bibinfo  {journal} {Physical Review E}\ }\textbf
  {\bibinfo {volume} {96}},\ \bibinfo {pages} {012211} (\bibinfo {year}
  {2017}{\natexlab{b}})}\BibitemShut {NoStop}%
\bibitem [{\citenamefont {Koppenh{\"o}fer}\ and\ \citenamefont
  {Roulet}(2019)}]{koppenhofer2019optimal}%
  \BibitemOpen
  \bibfield  {author} {\bibinfo {author} {\bibfnamefont {M.}~\bibnamefont
  {Koppenh{\"o}fer}}\ and\ \bibinfo {author} {\bibfnamefont {A.}~\bibnamefont
  {Roulet}},\ }\bibfield  {title} {\bibinfo {title} {Optimal synchronization
  deep in the quantum regime: Resource and fundamental limit},\ }\href
  {https://doi.org/10.1103/PhysRevA.99.043804} {\bibfield  {journal} {\bibinfo
  {journal} {Physical Review A}\ }\textbf {\bibinfo {volume} {99}},\ \bibinfo
  {pages} {043804} (\bibinfo {year} {2019})}\BibitemShut {NoStop}%
\bibitem [{\citenamefont {Mari}\ \emph {et~al.}(2013)\citenamefont {Mari},
  \citenamefont {Farace}, \citenamefont {Didier}, \citenamefont {Giovannetti},\
  and\ \citenamefont {Fazio}}]{mari2013prl}%
  \BibitemOpen
  \bibfield  {author} {\bibinfo {author} {\bibfnamefont {A.}~\bibnamefont
  {Mari}}, \bibinfo {author} {\bibfnamefont {A.}~\bibnamefont {Farace}},
  \bibinfo {author} {\bibfnamefont {N.}~\bibnamefont {Didier}}, \bibinfo
  {author} {\bibfnamefont {V.}~\bibnamefont {Giovannetti}},\ and\ \bibinfo
  {author} {\bibfnamefont {R.}~\bibnamefont {Fazio}},\ }\bibfield  {title}
  {\bibinfo {title} {Measures of quantum synchronization in continuous variable
  systems},\ }\href {https://doi.org/10.1103/PhysRevLett.111.103605} {\bibfield
   {journal} {\bibinfo  {journal} {Phys. Rev. Lett.}\ }\textbf {\bibinfo
  {volume} {111}},\ \bibinfo {pages} {103605} (\bibinfo {year}
  {2013})}\BibitemShut {NoStop}%
\bibitem [{\citenamefont {Jaseem}\ \emph
  {et~al.}(2020{\natexlab{b}})\citenamefont {Jaseem}, \citenamefont
  {Hajdu{\v{s}}ek}, \citenamefont {Solanki}, \citenamefont {Kwek},
  \citenamefont {Fazio},\ and\ \citenamefont
  {Vinjanampathy}}]{jaseem2020generalized}%
  \BibitemOpen
  \bibfield  {author} {\bibinfo {author} {\bibfnamefont {N.}~\bibnamefont
  {Jaseem}}, \bibinfo {author} {\bibfnamefont {M.}~\bibnamefont
  {Hajdu{\v{s}}ek}}, \bibinfo {author} {\bibfnamefont {P.}~\bibnamefont
  {Solanki}}, \bibinfo {author} {\bibfnamefont {L.-C.}\ \bibnamefont {Kwek}},
  \bibinfo {author} {\bibfnamefont {R.}~\bibnamefont {Fazio}},\ and\ \bibinfo
  {author} {\bibfnamefont {S.}~\bibnamefont {Vinjanampathy}},\ }\bibfield
  {title} {\bibinfo {title} {Generalized measure of quantum synchronization},\
  }\href {https://doi.org/10.1103/PhysRevResearch.2.043287} {\bibfield
  {journal} {\bibinfo  {journal} {Physical Review Research}\ }\textbf {\bibinfo
  {volume} {2}},\ \bibinfo {pages} {043287} (\bibinfo {year}
  {2020}{\natexlab{b}})}\BibitemShut {NoStop}%
\bibitem [{\citenamefont {Manzano}\ \emph {et~al.}(2013)\citenamefont
  {Manzano}, \citenamefont {Galve}, \citenamefont {Giorgi}, \citenamefont
  {Hern{\'a}ndez-Garc{\'\i}a},\ and\ \citenamefont
  {Zambrini}}]{manzano2013synchronization}%
  \BibitemOpen
  \bibfield  {author} {\bibinfo {author} {\bibfnamefont {G.}~\bibnamefont
  {Manzano}}, \bibinfo {author} {\bibfnamefont {F.}~\bibnamefont {Galve}},
  \bibinfo {author} {\bibfnamefont {G.~L.}\ \bibnamefont {Giorgi}}, \bibinfo
  {author} {\bibfnamefont {E.}~\bibnamefont {Hern{\'a}ndez-Garc{\'\i}a}},\ and\
  \bibinfo {author} {\bibfnamefont {R.}~\bibnamefont {Zambrini}},\ }\bibfield
  {title} {\bibinfo {title} {Synchronization, quantum correlations and
  entanglement in oscillator networks},\ }\href
  {https://doi.org/https://doi.org/10.1038/srep01439} {\bibfield  {journal}
  {\bibinfo  {journal} {Scientific Reports}\ }\textbf {\bibinfo {volume} {3}},\
  \bibinfo {pages} {1} (\bibinfo {year} {2013})}\BibitemShut {NoStop}%
\bibitem [{\citenamefont {Ameri}\ \emph {et~al.}(2015)\citenamefont {Ameri},
  \citenamefont {Eghbali-Arani}, \citenamefont {Mari}, \citenamefont {Farace},
  \citenamefont {Kheirandish}, \citenamefont {Giovannetti},\ and\ \citenamefont
  {Fazio}}]{ameri2015mutual}%
  \BibitemOpen
  \bibfield  {author} {\bibinfo {author} {\bibfnamefont {V.}~\bibnamefont
  {Ameri}}, \bibinfo {author} {\bibfnamefont {M.}~\bibnamefont
  {Eghbali-Arani}}, \bibinfo {author} {\bibfnamefont {A.}~\bibnamefont {Mari}},
  \bibinfo {author} {\bibfnamefont {A.}~\bibnamefont {Farace}}, \bibinfo
  {author} {\bibfnamefont {F.}~\bibnamefont {Kheirandish}}, \bibinfo {author}
  {\bibfnamefont {V.}~\bibnamefont {Giovannetti}},\ and\ \bibinfo {author}
  {\bibfnamefont {R.}~\bibnamefont {Fazio}},\ }\bibfield  {title} {\bibinfo
  {title} {Mutual information as an order parameter for quantum
  synchronization},\ }\href {https://doi.org/10.1103/PhysRevA.91.012301}
  {\bibfield  {journal} {\bibinfo  {journal} {Physical Review A}\ }\textbf
  {\bibinfo {volume} {91}},\ \bibinfo {pages} {012301} (\bibinfo {year}
  {2015})}\BibitemShut {NoStop}%
\bibitem [{\citenamefont {Roulet}\ and\ \citenamefont
  {Bruder}(2018{\natexlab{b}})}]{bruder2018smallest}%
  \BibitemOpen
  \bibfield  {author} {\bibinfo {author} {\bibfnamefont {A.}~\bibnamefont
  {Roulet}}\ and\ \bibinfo {author} {\bibfnamefont {C.}~\bibnamefont
  {Bruder}},\ }\bibfield  {title} {\bibinfo {title} {Synchronizing the smallest
  possible system},\ }\href {https://doi.org/10.1103/PhysRevLett.121.053601}
  {\bibfield  {journal} {\bibinfo  {journal} {Phys. Rev. Lett.}\ }\textbf
  {\bibinfo {volume} {121}},\ \bibinfo {pages} {053601} (\bibinfo {year}
  {2018}{\natexlab{b}})}\BibitemShut {NoStop}%
\bibitem [{\citenamefont {Sonar}\ \emph {et~al.}(2018)\citenamefont {Sonar},
  \citenamefont {Hajdu\ifmmode~\check{s}\else \v{s}\fi{}ek}, \citenamefont
  {Mukherjee}, \citenamefont {Fazio}, \citenamefont {Vedral}, \citenamefont
  {Vinjanampathy},\ and\ \citenamefont {Kwek}}]{sameer2018squuezing}%
  \BibitemOpen
  \bibfield  {author} {\bibinfo {author} {\bibfnamefont {S.}~\bibnamefont
  {Sonar}}, \bibinfo {author} {\bibfnamefont {M.}~\bibnamefont
  {Hajdu\ifmmode~\check{s}\else \v{s}\fi{}ek}}, \bibinfo {author}
  {\bibfnamefont {M.}~\bibnamefont {Mukherjee}}, \bibinfo {author}
  {\bibfnamefont {R.}~\bibnamefont {Fazio}}, \bibinfo {author} {\bibfnamefont
  {V.}~\bibnamefont {Vedral}}, \bibinfo {author} {\bibfnamefont
  {S.}~\bibnamefont {Vinjanampathy}},\ and\ \bibinfo {author} {\bibfnamefont
  {L.-C.}\ \bibnamefont {Kwek}},\ }\bibfield  {title} {\bibinfo {title}
  {Squeezing enhances quantum synchronization},\ }\href
  {https://doi.org/10.1103/PhysRevLett.120.163601} {\bibfield  {journal}
  {\bibinfo  {journal} {Phys. Rev. Lett.}\ }\textbf {\bibinfo {volume} {120}},\
  \bibinfo {pages} {163601} (\bibinfo {year} {2018})}\BibitemShut {NoStop}%
\bibitem [{\citenamefont {Buca}\ \emph {et~al.}(2021)\citenamefont {Buca},
  \citenamefont {Booker},\ and\ \citenamefont {Jaksch}}]{buca2021algebraic}%
  \BibitemOpen
  \bibfield  {author} {\bibinfo {author} {\bibfnamefont {B.}~\bibnamefont
  {Buca}}, \bibinfo {author} {\bibfnamefont {C.}~\bibnamefont {Booker}},\ and\
  \bibinfo {author} {\bibfnamefont {D.}~\bibnamefont {Jaksch}},\ }\bibfield
  {title} {\bibinfo {title} {Algebraic theory of quantum synchronization and
  limit cycles under dissipation},\ }\href@noop {} {\bibfield  {journal}
  {\bibinfo  {journal} {arXiv preprint arXiv:2103.01808}\ } (\bibinfo {year}
  {2021})}\BibitemShut {NoStop}%
\bibitem [{\citenamefont {Koppenh{\"o}fer}\ \emph {et~al.}(2020)\citenamefont
  {Koppenh{\"o}fer}, \citenamefont {Bruder},\ and\ \citenamefont
  {Roulet}}]{koppenhofer2020quantum}%
  \BibitemOpen
  \bibfield  {author} {\bibinfo {author} {\bibfnamefont {M.}~\bibnamefont
  {Koppenh{\"o}fer}}, \bibinfo {author} {\bibfnamefont {C.}~\bibnamefont
  {Bruder}},\ and\ \bibinfo {author} {\bibfnamefont {A.}~\bibnamefont
  {Roulet}},\ }\bibfield  {title} {\bibinfo {title} {Quantum synchronization on
  the ibm q system},\ }\href {https://doi.org/10.1103/PhysRevResearch.2.023026}
  {\bibfield  {journal} {\bibinfo  {journal} {Physical Review Research}\
  }\textbf {\bibinfo {volume} {2}},\ \bibinfo {pages} {023026} (\bibinfo {year}
  {2020})}\BibitemShut {NoStop}%
\bibitem [{\citenamefont {Laskar}\ \emph {et~al.}(2020)\citenamefont {Laskar},
  \citenamefont {Adhikary}, \citenamefont {Mondal}, \citenamefont {Katiyar},
  \citenamefont {Vinjanampathy},\ and\ \citenamefont
  {Ghosh}}]{laskar2020observation}%
  \BibitemOpen
  \bibfield  {author} {\bibinfo {author} {\bibfnamefont {A.~W.}\ \bibnamefont
  {Laskar}}, \bibinfo {author} {\bibfnamefont {P.}~\bibnamefont {Adhikary}},
  \bibinfo {author} {\bibfnamefont {S.}~\bibnamefont {Mondal}}, \bibinfo
  {author} {\bibfnamefont {P.}~\bibnamefont {Katiyar}}, \bibinfo {author}
  {\bibfnamefont {S.}~\bibnamefont {Vinjanampathy}},\ and\ \bibinfo {author}
  {\bibfnamefont {S.}~\bibnamefont {Ghosh}},\ }\bibfield  {title} {\bibinfo
  {title} {Observation of quantum phase synchronization in spin-1 atoms},\
  }\href {https://doi.org/10.1103/PhysRevLett.125.013601} {\bibfield  {journal}
  {\bibinfo  {journal} {Physical Review Letters}\ }\textbf {\bibinfo {volume}
  {125}},\ \bibinfo {pages} {013601} (\bibinfo {year} {2020})}\BibitemShut
  {NoStop}%
\bibitem [{\citenamefont {Bengtsson}\ and\ \citenamefont
  {Zyczkowski}(2006)}]{bengtsson2006geometry}%
  \BibitemOpen
  \bibfield  {author} {\bibinfo {author} {\bibfnamefont {I.}~\bibnamefont
  {Bengtsson}}\ and\ \bibinfo {author} {\bibfnamefont {K.}~\bibnamefont
  {Zyczkowski}},\ }\href
  {https://doi.org/https://doi.org/10.1017/CBO9780511535048} {\emph {\bibinfo
  {title} {Geometry of Qauntum States: An Introduction to Quantum
  Entanglement}}}\ (\bibinfo  {publisher} {Cambridge University Press},\
  \bibinfo {year} {2006})\BibitemShut {NoStop}%
\bibitem [{\citenamefont {Krithika}\ \emph {et~al.}(2019)\citenamefont
  {Krithika}, \citenamefont {Anjusha}, \citenamefont {Bhosale},\ and\
  \citenamefont {Mahesh}}]{krithika2019nmr}%
  \BibitemOpen
  \bibfield  {author} {\bibinfo {author} {\bibfnamefont {V.~R.}\ \bibnamefont
  {Krithika}}, \bibinfo {author} {\bibfnamefont {V.~S.}\ \bibnamefont
  {Anjusha}}, \bibinfo {author} {\bibfnamefont {U.~T.}\ \bibnamefont
  {Bhosale}},\ and\ \bibinfo {author} {\bibfnamefont {T.~S.}\ \bibnamefont
  {Mahesh}},\ }\bibfield  {title} {\bibinfo {title} {Nmr studies of quantum
  chaos in a two-qubit kicked top},\ }\href
  {https://doi.org/10.1103/PhysRevE.99.032219} {\bibfield  {journal} {\bibinfo
  {journal} {Physical Review E}\ }\textbf {\bibinfo {volume} {99}},\ \bibinfo
  {pages} {032219} (\bibinfo {year} {2019})}\BibitemShut {NoStop}%
\bibitem [{\citenamefont {Husimi}(1940)}]{husimi1940some}%
  \BibitemOpen
  \bibfield  {author} {\bibinfo {author} {\bibfnamefont {K.}~\bibnamefont
  {Husimi}},\ }\bibfield  {title} {\bibinfo {title} {Some formal properties of
  the density matrix},\ }\href {https://doi.org/10.11429/ppmsj1919.22.4_264}
  {\bibfield  {journal} {\bibinfo  {journal} {Proceedings of the
  Physico-Mathematical Society of Japan. 3rd Series}\ }\textbf {\bibinfo
  {volume} {22}},\ \bibinfo {pages} {264} (\bibinfo {year} {1940})}\BibitemShut
  {NoStop}%
\bibitem [{\citenamefont {Kano}(1965)}]{kano1965new}%
  \BibitemOpen
  \bibfield  {author} {\bibinfo {author} {\bibfnamefont {Y.}~\bibnamefont
  {Kano}},\ }\bibfield  {title} {\bibinfo {title} {A new phase-space
  distribution function in the statistical theory of the electromagnetic
  field},\ }\href {https://doi.org/10.1063/1.1704739} {\bibfield  {journal}
  {\bibinfo  {journal} {Journal of Mathematical Physics}\ }\textbf {\bibinfo
  {volume} {6}},\ \bibinfo {pages} {1913} (\bibinfo {year} {1965})}\BibitemShut
  {NoStop}%
\bibitem [{\citenamefont {Perelomov}(1986)}]{perelomov1986}%
  \BibitemOpen
  \bibfield  {author} {\bibinfo {author} {\bibfnamefont {A.}~\bibnamefont
  {Perelomov}},\ }\href
  {https://doi.org/https://doi.org/10.1007/978-3-642-61629-7} {\emph {\bibinfo
  {title} {Generalized Coherent States and Their Applications}}}\ (\bibinfo
  {publisher} {Springer, Berlin, Heidelberg},\ \bibinfo {year}
  {1986})\BibitemShut {NoStop}%
\bibitem [{\citenamefont {Nemoto}(2000)}]{nemoto2000generalized}%
  \BibitemOpen
  \bibfield  {author} {\bibinfo {author} {\bibfnamefont {K.}~\bibnamefont
  {Nemoto}},\ }\bibfield  {title} {\bibinfo {title} {Generalized coherent
  states for su (n) systems},\ }\href
  {https://doi.org/10.1088/0305-4470/33/17/307} {\bibfield  {journal} {\bibinfo
   {journal} {Journal of Physics A: Mathematical and General}\ }\textbf
  {\bibinfo {volume} {33}},\ \bibinfo {pages} {3493} (\bibinfo {year}
  {2000})}\BibitemShut {NoStop}%
\bibitem [{\citenamefont {Mathur}\ and\ \citenamefont
  {Mani}(2002)}]{mathur2002n}%
  \BibitemOpen
  \bibfield  {author} {\bibinfo {author} {\bibfnamefont {M.}~\bibnamefont
  {Mathur}}\ and\ \bibinfo {author} {\bibfnamefont {H.}~\bibnamefont {Mani}},\
  }\bibfield  {title} {\bibinfo {title} {Su (n) coherent states},\ }\href
  {https://doi.org/10.1063/1.1513651} {\bibfield  {journal} {\bibinfo
  {journal} {Journal of Mathematical Physics}\ }\textbf {\bibinfo {volume}
  {43}},\ \bibinfo {pages} {5351} (\bibinfo {year} {2002})}\BibitemShut
  {NoStop}%
\bibitem [{\citenamefont {Breuer}\ \emph {et~al.}(2002)\citenamefont {Breuer},
  \citenamefont {Petruccione} \emph {et~al.}}]{breuer2002theory}%
  \BibitemOpen
  \bibfield  {author} {\bibinfo {author} {\bibfnamefont {H.-P.}\ \bibnamefont
  {Breuer}}, \bibinfo {author} {\bibfnamefont {F.}~\bibnamefont {Petruccione}},
  \emph {et~al.},\ }\href
  {https://doi.org/10.1093/acprof:oso/9780199213900.001.0001} {\emph {\bibinfo
  {title} {The theory of open quantum systems}}}\ (\bibinfo  {publisher}
  {Oxford University Press on Demand},\ \bibinfo {year} {2002})\BibitemShut
  {NoStop}%
\bibitem [{\citenamefont {Ghosh}\ \emph {et~al.}(2012)\citenamefont {Ghosh},
  \citenamefont {Sinha},\ and\ \citenamefont {Ray}}]{ghosh2012fermionic}%
  \BibitemOpen
  \bibfield  {author} {\bibinfo {author} {\bibfnamefont {A.}~\bibnamefont
  {Ghosh}}, \bibinfo {author} {\bibfnamefont {S.~S.}\ \bibnamefont {Sinha}},\
  and\ \bibinfo {author} {\bibfnamefont {D.~S.}\ \bibnamefont {Ray}},\
  }\bibfield  {title} {\bibinfo {title} {Fermionic oscillator in a fermionic
  bath},\ }\href {https://doi.org/10.1103/PhysRevE.86.011138} {\bibfield
  {journal} {\bibinfo  {journal} {Physical Review E}\ }\textbf {\bibinfo
  {volume} {86}},\ \bibinfo {pages} {011138} (\bibinfo {year}
  {2012})}\BibitemShut {NoStop}%
\bibitem [{\citenamefont {Suri}\ \emph {et~al.}(2018)\citenamefont {Suri},
  \citenamefont {Binder}, \citenamefont {Muralidharan},\ and\ \citenamefont
  {Vinjanampathy}}]{suri2018speeding}%
  \BibitemOpen
  \bibfield  {author} {\bibinfo {author} {\bibfnamefont {N.}~\bibnamefont
  {Suri}}, \bibinfo {author} {\bibfnamefont {F.~C.}\ \bibnamefont {Binder}},
  \bibinfo {author} {\bibfnamefont {B.}~\bibnamefont {Muralidharan}},\ and\
  \bibinfo {author} {\bibfnamefont {S.}~\bibnamefont {Vinjanampathy}},\
  }\bibfield  {title} {\bibinfo {title} {Speeding up thermalisation via open
  quantum system variational optimisation},\ }\href
  {https://doi.org/https://doi.org/10.1140/epjst/e2018-00125-6} {\bibfield
  {journal} {\bibinfo  {journal} {The European Physical Journal Special
  Topics}\ }\textbf {\bibinfo {volume} {227}},\ \bibinfo {pages} {203}
  (\bibinfo {year} {2018})}\BibitemShut {NoStop}%
\bibitem [{\citenamefont {Manzano}(2020)}]{manzano2020short}%
  \BibitemOpen
  \bibfield  {author} {\bibinfo {author} {\bibfnamefont {D.}~\bibnamefont
  {Manzano}},\ }\bibfield  {title} {\bibinfo {title} {A short introduction to
  the lindblad master equation},\ }\href {https://doi.org/10.1063/1.5115323}
  {\bibfield  {journal} {\bibinfo  {journal} {AIP Advances}\ }\textbf {\bibinfo
  {volume} {10}},\ \bibinfo {pages} {025106} (\bibinfo {year}
  {2020})}\BibitemShut {NoStop}%
\bibitem [{\citenamefont {Gyamfi}(2020)}]{gyamfi2020fundamentals}%
  \BibitemOpen
  \bibfield  {author} {\bibinfo {author} {\bibfnamefont {J.~A.}\ \bibnamefont
  {Gyamfi}},\ }\bibfield  {title} {\bibinfo {title} {Fundamentals of quantum
  mechanics in liouville space},\ }\href
  {https://doi.org/10.1088/1361-6404/ab9fdd} {\bibfield  {journal} {\bibinfo
  {journal} {European Journal of Physics}\ }\textbf {\bibinfo {volume} {41}},\
  \bibinfo {pages} {063002} (\bibinfo {year} {2020})}\BibitemShut {NoStop}%
\bibitem [{\citenamefont {Navarrete-Benlloch}(2015)}]{navarrete2015open}%
  \BibitemOpen
  \bibfield  {author} {\bibinfo {author} {\bibfnamefont {C.}~\bibnamefont
  {Navarrete-Benlloch}},\ }\bibfield  {title} {\bibinfo {title} {Open systems
  dynamics: Simulating master equations in the computer},\ }\href@noop {}
  {\bibfield  {journal} {\bibinfo  {journal} {arXiv preprint arXiv:1504.05266}\
  } (\bibinfo {year} {2015})}\BibitemShut {NoStop}%
\bibitem [{\citenamefont {Boulant}\ \emph {et~al.}(2003)\citenamefont
  {Boulant}, \citenamefont {Havel}, \citenamefont {Pravia},\ and\ \citenamefont
  {Cory}}]{boulant2003robust}%
  \BibitemOpen
  \bibfield  {author} {\bibinfo {author} {\bibfnamefont {N.}~\bibnamefont
  {Boulant}}, \bibinfo {author} {\bibfnamefont {T.~F.}\ \bibnamefont {Havel}},
  \bibinfo {author} {\bibfnamefont {M.~A.}\ \bibnamefont {Pravia}},\ and\
  \bibinfo {author} {\bibfnamefont {D.~G.}\ \bibnamefont {Cory}},\ }\bibfield
  {title} {\bibinfo {title} {Robust method for estimating the lindblad
  operators of a dissipative quantum process from measurements of the density
  operator at multiple time points},\ }\href
  {https://doi.org/10.1103/PhysRevA.67.042322} {\bibfield  {journal} {\bibinfo
  {journal} {Physical Review A}\ }\textbf {\bibinfo {volume} {67}},\ \bibinfo
  {pages} {042322} (\bibinfo {year} {2003})}\BibitemShut {NoStop}%
\bibitem [{\citenamefont {Chuang}\ \emph {et~al.}(1998)\citenamefont {Chuang},
  \citenamefont {Gershenfeld}, \citenamefont {Kubinec},\ and\ \citenamefont
  {Leung}}]{chuang1998bulk}%
  \BibitemOpen
  \bibfield  {author} {\bibinfo {author} {\bibfnamefont {I.~L.}\ \bibnamefont
  {Chuang}}, \bibinfo {author} {\bibfnamefont {N.}~\bibnamefont {Gershenfeld}},
  \bibinfo {author} {\bibfnamefont {M.~G.}\ \bibnamefont {Kubinec}},\ and\
  \bibinfo {author} {\bibfnamefont {D.~W.}\ \bibnamefont {Leung}},\ }\bibfield
  {title} {\bibinfo {title} {Bulk quantum computation with nuclear magnetic
  resonance: theory and experiment},\ }\href
  {https://doi.org/https://doi.org/10.1098/rspa.1998.0170} {\bibfield
  {journal} {\bibinfo  {journal} {Proceedings of the Royal Society of London.
  Series A: Mathematical, Physical and Engineering Sciences}\ }\textbf
  {\bibinfo {volume} {454}},\ \bibinfo {pages} {447} (\bibinfo {year}
  {1998})}\BibitemShut {NoStop}%
\bibitem [{\citenamefont {Shukla}\ \emph {et~al.}(2013)\citenamefont {Shukla},
  \citenamefont {Rao},\ and\ \citenamefont {Mahesh}}]{shukla2013ancilla}%
  \BibitemOpen
  \bibfield  {author} {\bibinfo {author} {\bibfnamefont {A.}~\bibnamefont
  {Shukla}}, \bibinfo {author} {\bibfnamefont {K.~R.~K.}\ \bibnamefont {Rao}},\
  and\ \bibinfo {author} {\bibfnamefont {T.~S.}\ \bibnamefont {Mahesh}},\
  }\bibfield  {title} {\bibinfo {title} {Ancilla-assisted quantum state
  tomography in multiqubit registers},\ }\href
  {https://doi.org/10.1103/PhysRevA.87.062317} {\bibfield  {journal} {\bibinfo
  {journal} {Physical Review A}\ }\textbf {\bibinfo {volume} {87}},\ \bibinfo
  {pages} {062317} (\bibinfo {year} {2013})}\BibitemShut {NoStop}%
\bibitem [{\citenamefont {Lee~Loh}\ and\ \citenamefont
  {Kim}(2015)}]{lee2015visualizing}%
  \BibitemOpen
  \bibfield  {author} {\bibinfo {author} {\bibfnamefont {Y.}~\bibnamefont
  {Lee~Loh}}\ and\ \bibinfo {author} {\bibfnamefont {M.}~\bibnamefont {Kim}},\
  }\bibfield  {title} {\bibinfo {title} {Visualizing spin states using the spin
  coherent state representation},\ }\href {https://doi.org/10.1119/1.4898595}
  {\bibfield  {journal} {\bibinfo  {journal} {American Journal of Physics}\
  }\textbf {\bibinfo {volume} {83}},\ \bibinfo {pages} {30} (\bibinfo {year}
  {2015})}\BibitemShut {NoStop}%
\bibitem [{\citenamefont {Radcliffe}(1971)}]{radcliffe1971some}%
  \BibitemOpen
  \bibfield  {author} {\bibinfo {author} {\bibfnamefont {J.}~\bibnamefont
  {Radcliffe}},\ }\bibfield  {title} {\bibinfo {title} {Some properties of spin
  coherent states},\ }\href {https://doi.org/10.1088/0305-4470/4/3/009}
  {\bibfield  {journal} {\bibinfo  {journal} {Phys. A: Math. Gen}\ }\textbf
  {\bibinfo {volume} {4}},\ \bibinfo {pages} {313} (\bibinfo {year}
  {1971})}\BibitemShut {NoStop}%
\bibitem [{\citenamefont {Klauder}\ and\ \citenamefont
  {Skagerstam}(1985)}]{klauder1985generalized}%
  \BibitemOpen
  \bibfield  {author} {\bibinfo {author} {\bibfnamefont {J.~R.}\ \bibnamefont
  {Klauder}}\ and\ \bibinfo {author} {\bibfnamefont {B.}~\bibnamefont
  {Skagerstam}},\ }\href {https://doi.org/10.1142/0096} {\emph {\bibinfo
  {title} {Coherent Sattes:Applications in Physics and Mathematical Physics}}}\
  (\bibinfo  {publisher} {World Scientific},\ \bibinfo {year}
  {1985})\BibitemShut {NoStop}%
\bibitem [{\citenamefont {Pande}\ \emph {et~al.}(2017)\citenamefont {Pande},
  \citenamefont {Bhole}, \citenamefont {Khurana},\ and\ \citenamefont
  {Mahesh}}]{VaradPRA}%
  \BibitemOpen
  \bibfield  {author} {\bibinfo {author} {\bibfnamefont {V.~R.}\ \bibnamefont
  {Pande}}, \bibinfo {author} {\bibfnamefont {G.}~\bibnamefont {Bhole}},
  \bibinfo {author} {\bibfnamefont {D.}~\bibnamefont {Khurana}},\ and\ \bibinfo
  {author} {\bibfnamefont {T.~S.}\ \bibnamefont {Mahesh}},\ }\bibfield  {title}
  {\bibinfo {title} {Strong algorithmic cooling in large star-topology quantum
  registers},\ }\href {https://doi.org/10.1103/PhysRevA.96.012330} {\bibfield
  {journal} {\bibinfo  {journal} {Phys. Rev. A}\ }\textbf {\bibinfo {volume}
  {96}},\ \bibinfo {pages} {012330} (\bibinfo {year} {2017})}\BibitemShut
  {NoStop}%
\end{thebibliography}%
\end{document}